\documentclass{jfm}
\usepackage{graphicx}
\usepackage{epstopdf, epsfig}
\usepackage{multirow}
\usepackage{mathrsfs,amsmath}
\usepackage{bm}
\usepackage{amssymb}
\usepackage{multirow,bigdelim}
\usepackage{upgreek}
\usepackage{units}
\usepackage{array,multirow}
\usepackage{booktabs}
\usepackage{psfrag}
\usepackage{tikz}
\usetikzlibrary{arrows.meta}
\usepackage{color}
\usepackage{xcolor}
\usepackage{threeparttable}	
\usepackage{dcolumn}		
\newcolumntype{d}[1]{D{.}{.}{#1}}
\usetikzlibrary{decorations}
\usetikzlibrary{decorations.markings}
\usetikzlibrary{calc,decorations.pathreplacing}
\usepackage{lipsum}
\usepackage{mathtools}
\usepackage{ar}
\usepackage{enumitem}
\usepackage{url}
\usepackage{wasysym}

\definecolor{Bblack}{rgb}{0.00, 0.00, 0.00}
\definecolor{Bblue}{rgb}{0,0.572,0.812}
\definecolor{Bgrey}{rgb}{0.25, 0.25, 0.25}

\newcommand*{\myfont}{\fontfamily{ptmri}\selectfont}
\DeclareTextFontCommand{\textmyfont}{\myfont}

\shorttitle{Decomposition of the streamwise TKE in boundary layers. Part 1}
\shortauthor{W. J. Baars and I. Marusic}
\title{Data-driven decomposition of the streamwise turbulence kinetic energy in boundary layers. Part 1. Energy spectra}

\author{Woutijn J. Baars$^1$
  \corresp{\email{baars@eng.au.dk}}
  \and
  Ivan Marusic$^2$}

\affiliation{$^1$Department of Engineering, Aarhus University, 8000 Aarhus C, Denmark \\[\affilskip] $^2$Department of Mechanical Engineering, The University of Melbourne, VIC 3010, Australia}

\begin{document}

\maketitle
\begin{abstract}
In wall-bounded turbulence, a multitude of coexisting turbulence structures form the streamwise velocity energy spectrum from the viscosity- to the inertia-dominated range of scales. Definite scaling-trends for streamwise spectra have remained empirically-elusive, although a prominent school of thought stems from the works of \citet*{perry:1977a} and \citet*{perry:1986a}, which were greatly inspired by the attached-eddy hypothesis of \citet*{townsend:1976bk}. In this paper, we re-examine the turbulence kinetic energy of the streamwise velocity component in the context of the spectral decompositions of Perry and coworkers. Two universal spectral filters are derived via spectral coherence analysis of two-point velocity signals, spanning a Reynolds number range $Re_\tau \sim \mathcal{O}(10^3)$ to $\mathcal{O}(10^6)$ and form the basis for our decomposition of the logarithmic-region turbulence into stochastically wall-detached and wall-attached portions of energy. The latter is composed of scales larger than a streamwise/wall-normal ratio of $\lambda_x/z \approx 14$. If the decomposition is accepted, a $k_x^{-1}$ scaling region can only appear for $Re_\tau \apprge 80\,000$, at a wall-normal position of $z^+ = 100$. Following Perry and co-workers, it is hypothesized that spectral contributions from turbulence structures other than attached eddies obscure a $k_x^{-1}$ scaling. When accepting the idea of different spectral contributions it is furthermore shown that a broad outer-spectral peak is present even at low $Re_\tau$.
\end{abstract}

\section{Introduction and context}\label{sec:intro}
A long-standing challenge in the study of wall turbulence has been the scalings of energy spectra, amongst other statistical quantities. For the fluctuations of the streamwise velocity component $u$, the energy spectrum is denoted as $\phi_{uu}(k_x)$, with $k_x$ being the streamwise wavenumber. The streamwise turbulence intensity refers to the velocity variance, $\overline{u^2}$, and equates to the integrated spectral energy: $\overline{u^2} = \int \phi_{uu} {\rm d}k_x$. Thus, energy spectra inform how the turbulence intensity is distributed across wavenumbers and have long been used to interpret the turbulence cascade \citep[\emph{e.g.}][]{pope:2000bk,jimenez:2012a}. For wall-bounded turbulence, the inhomogeneity along its wall-normal direction and the anisotropy of the inertial motions, introduce mathematical challenges that prevent derivations of spectral scaling laws for the full spectrum. In addition, research over the past two decades has revealed large-scale organized motions with significant spatial and temporal coherence \citep{robinson:1991a,adrian:2007a,hutchins:2007aa,smits:2011a,jimenez:2018a}. In this paper we consider $\phi_{uu}$ corresponding to zero-pressure gradient (ZPG) turbulent boundary layers (TBLs). A central aspect is a data-driven spectral decomposition, yielding a spectral sub-component that resonates with one that would be induced by eddies obeying Townsend's attached-eddy hypothesis (AEH) \citep{townsend:1976bk,marusic:1995a,baidya:2017a}.

In \S\,\ref{sec:structure} we first review the current state of knowledge on the scaling of $\phi_{uu}$ and end with an outline of the paper. Throughout our work we use co-ordinate system $x$, $y$ and $z$ to denote the streamwise, spanwise and wall-normal directions of the flow, respectively. Reynolds number $Re_\tau \equiv \delta U_\tau/\nu$ is the ratio of $\delta$ (the boundary layer thickness) to the viscous length-scale $\nu/U_\tau$, where $\nu$ is the kinematic viscosity and $U_\tau$ is the friction velocity. Here $U_\tau = \sqrt{\tau_o/\rho}$, with $\tau_0$ and $\rho$ being the wall-shear stress and fluid's density, respectively. When a dimension of length is presented in \emph{outer-scaling}, it is normalized with $\delta$, while a \emph{viscous-scaling} is signified with superscript `+' and comprises a scaling with $\nu/U_\tau$. A \emph{wall-scaling} (or inner-scaling when spectra are concerned) employs wall-distance $z$. Recall that lower-case $u$ represents the Reynolds decomposed fluctuations, while capital $U$ is used for the absolute mean. We generally deal with time-series, $u(t)$, from which frequency spectra $\phi_{uu}(f)$ are computed ($f$ is the temporal frequency). For interpretative purposes and convention, frequency spectra are converted to wavenumber spectra using a single convection velocity $U_c$ (generally taken as the local mean velocity $U(z)$; wavenumber $k_x = 2\pi f/U_c$). Wavenumber spectra $\phi_{uu}(k_x)$ are computed as $\phi_{uu}(k_x) = \phi_{uu}(f)\left(df/dk\right)$, where factor $df/dk = U_c/(2\pi)$ converts the energy density from `per unit frequency' to `per unit wavenumber'. Note that premultiplying the spectrum, with $f$ or $k_x$, does not affect its magnitude. That is, $f\phi_{uu}(f) = 2\pi f/U_c \phi_{uu}(f) \left(df/dk\right) = k_x \phi_{uu}(k_x)$. The scale axis is either presented in terms of $k_x$ or wavelength $\lambda_x = 2\pi/k_x$. Frequency-to-wavenumber conversions are non-trivial and Taylor's hypothesis can introduce aliasing-type discrepancies, particularly in the near-wall region \citep{perry:1990a,delalamo:2009a,dekat:2015a,renard:2015a}. Conclusions in this article are made with regard to the logarithmic region, but caution in interpreting frequency spectra via length scales is noted.

\section{Structure of the streamwise turbulence kinetic energy}\label{sec:structure}
\subsection{Review of empirically observed trends for $\phi_{uu}$ energy spectra}\label{sec:revspectra}
\begin{figure} 
\vspace{10pt}
\centering
\includegraphics[width = 0.999\textwidth]{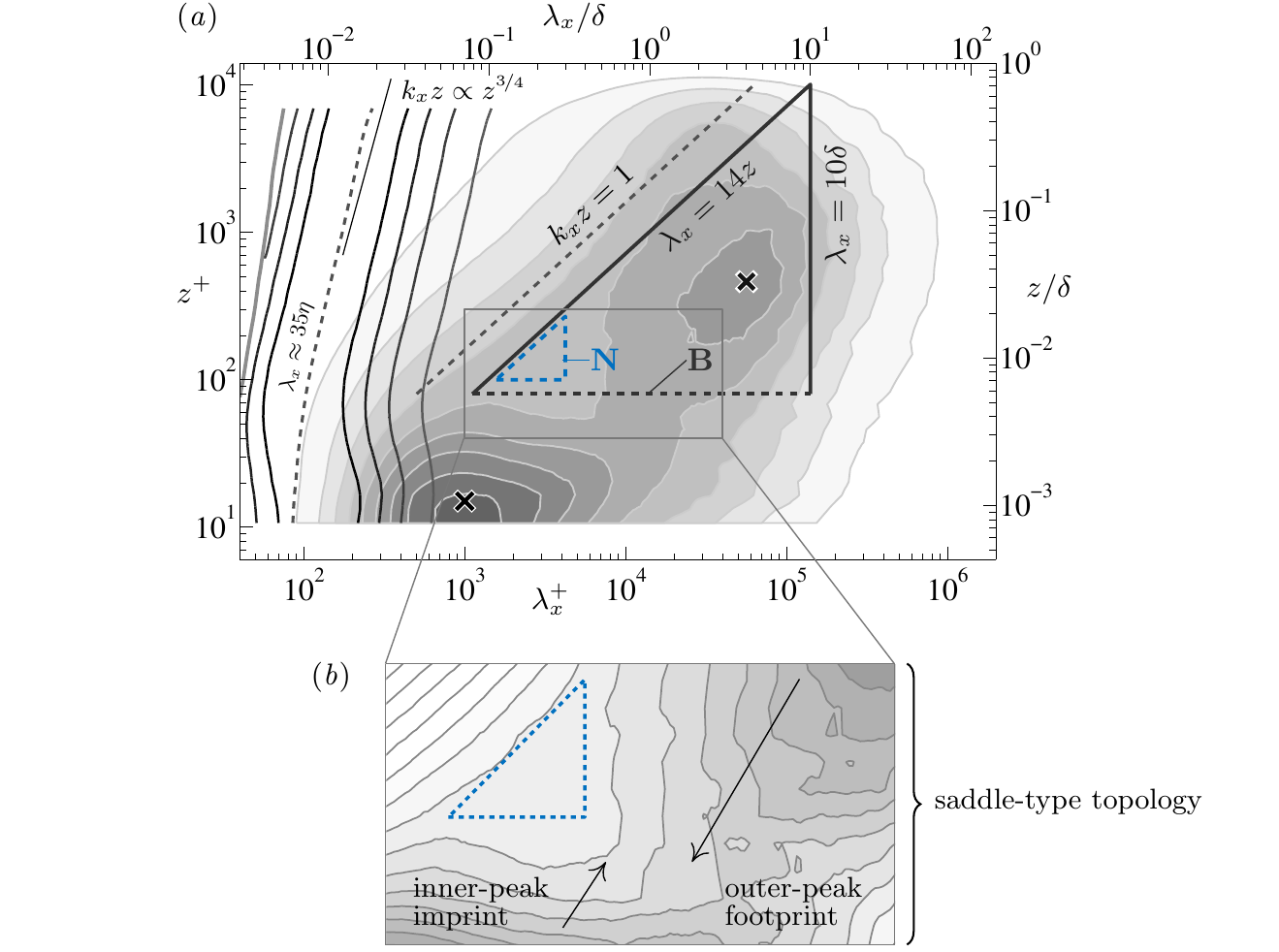}
   \caption{(\emph{a}) Premultiplied energy spectrogram $k^+_x\phi^+_{uu}$ (filled iso-contours 0.2:0.2:1.8) at $Re_\tau \approx 14\,100$ \citep{baars:2017a}, alongside a spectrogram of the surrogate dissipation $k^3_x\phi_{uu}$ (at each $z$, the dissipation spectrum is normalized by its maximum value; iso-contours 0.2:0.2:0.8). The dashed line tracks the ridge in the dissipation ($\lambda_x = 35\eta$). Triangle `N' refers to the $k_x^{-1}$ region identified by \citet{nickels:2005a}, while `B' labels a region where \citet{baars:2017a2} contemplated a self-similar behaviour. Note: $\lambda_x \equiv U(z)/f$. (\emph{b}) Saddle-type topology formed by the inner-peak imprint and outer-peak footprint.}
   \label{fig:spintro1}
\end{figure}

Over the past two decades, wall-turbulence data at high $Re_\tau$ have revealed new features of $\phi_{uu}$. Figure~\ref{fig:spintro1}(\emph{a}) presents the streamwise energy spectra as a spectrogram: premultiplied spectra at 40 logarithmically-spaced positions within the range $10.6 \apprle z^+ \apprle \delta^+$ are presented with iso-contours of $k^+_x\phi^+_{uu}$. These spectra were obtained from hot-wire measurements at $Re_\tau \approx 14\,100$ in Melbourne's TBL facility \citep{baars:2017a}.

Dominant small-scale features known as recurrent near-wall streaks \citep{kline:1967a} form the inner-spectral peak in the TBL spectrogram and obey viscous scaling (identified with the $\times$ marker at $\lambda_x^+ = 10^3$ and $z^+ = 15$ in Figure~\ref{fig:spintro1}\emph{a}). Attached eddies play an important role in the log-layer, and likely take a form consistent with vortex packets \citep{head:1981a,kim:1999a,adrian:2000a,wu:2006a,delalamo:2006a,balakumar:2007a,adrian:2007a}. These packets, or large-scale motions (LSMs), exhibit a forward inclined structure with $u < 0$ within the packet and $u > 0$ at either spanwise-flanked side of the packet \citep{favre:1967a,blackwelder:1972a,brown:1977a,wark:1991a,ganapathisubramani:2003a,tomkins:2003a}. Larger motions, known as superstructures or very large-scale motions (VLSMs), have also been described \citep{hutchins:2007ab}. In terms of their spectral signature, these large-scale organized motions appear most prominently as a broad spectral peak in the log-region, indicated in Figure~\ref{fig:spintro1}(\emph{a}) with a $\times$ marker at $\lambda_x = 4\delta$ and $z^+ = 3.9 Re_\tau^{\nicefrac{1}{2}} \approx 464$ \citep{mathis:2009a}. \citet{hutchins:2007aa} observed the emergence of the outer-spectral peak for $Re_\tau \gtrsim 2\,000$ and its energetic relevance increases with $Re_\tau$.
\begin{figure} 
\vspace{10pt}
\centering
\includegraphics[width = 0.999\textwidth]{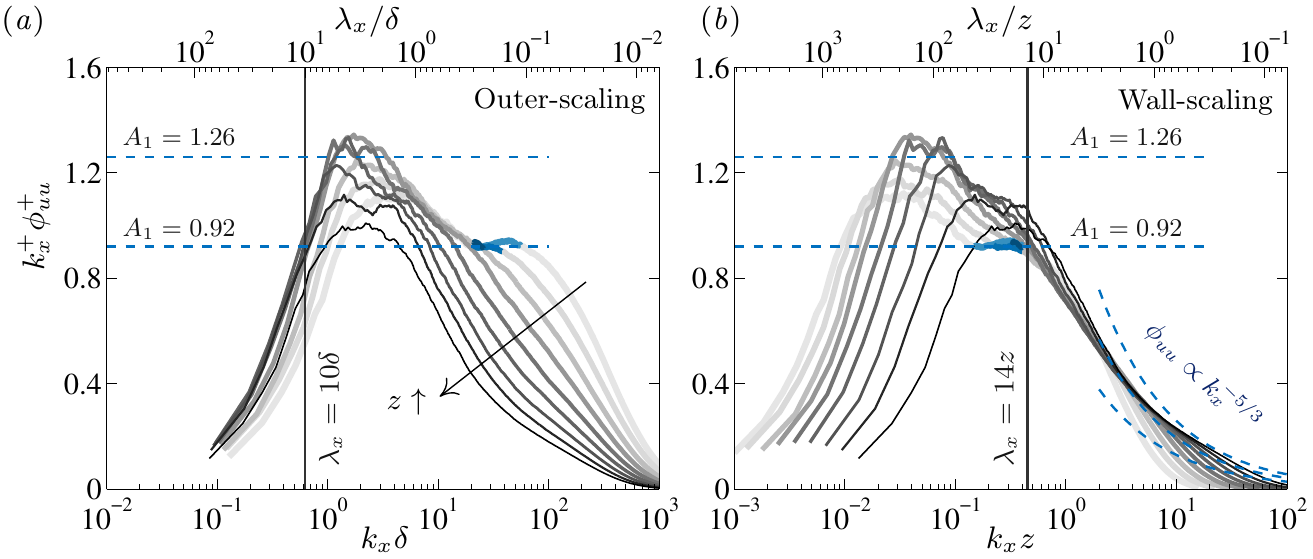}
   \caption{Premultiplied energy spectra ($Re_\tau \approx 14\,100$) at nine logarithmically-spaced wall-normal positions in the range $100 \apprle z^+ \apprle 0.15\delta^+$, with the abscissae in (a) outer-scaling and (b) wall-scaling, respectively. Note: $k_x \equiv 2\pi f/U(z)$.}
   \label{fig:spintro2}
\end{figure}

Figure~\ref{fig:spintro2} aids in assessing scalings of $\phi_{uu}$ in the logarithmic region. Individual spectra at logarithmically-spaced positions in the range $100 \apprle z^+ \apprle 0.15\delta^+$ are shown in Figure~\ref{fig:spintro2}(\emph{a}) with an outer-scaling on the logarithmic scale axis (and with a wall-scaling in Figure~\ref{fig:spintro2}\emph{b}). At the high wavenumber-end, spectra arguably resemble a $k_x^{-5/3}$ inertial sub-range scaling \citep[Figure~\ref{fig:spintro2}\emph{b} and][]{saddoughi:1994a,sreenivasan:1995a,mydlarsky:1996a,gamard:2000a,samie:2018a}. A decrease in spectral separation between the isotropic scales and the range over which viscosity acts, narrows the $k_x^{-5/3}$ region with decreasing $z$. In that regard, a second spectrogram in Figure~\ref{fig:spintro1}(\emph{a}) displays $k_x^3 \phi_{uu}$, indicative of the surrogate dissipation rate of energy $\epsilon \equiv 15\nu \int k_x^3\phi_{uu}{\rm d}\log(k_x)$ (with an applied local isotropy hypothesis, valid only well into the logarithmic region). In the logarithmic region, $\epsilon \sim 1/z$ and the dissipation scales via $k_xz \propto z^{\nicefrac{3}{4}}$ (production-dissipation balance); this scaling reflects the ridge of the dissipation spectrogram following $\lambda_x \approx 35\eta$, where $\eta \equiv (\nu^3/\epsilon)^{\nicefrac{1}{4}}$. When spectra adhere to the $k_x^{-5/3}$ up to a constant wall-scaling, \emph{e.g.} $k_x z = \mathcal{O}(1)$ (dashed line in Figure~\ref{fig:spintro1}\emph{a}), the dissipation and energy spectrograms reveal the relatively small range over which a $k_x^{-5/3}$ may be expected.

When concentrating on the larger scales, adjacent to the inertial sub-range, a combination of dimensional analysis, a spectral overlap argument and an assumed type of eddy similarity lead to a $k_x^{-1}$ dependence: \citep[\emph{e.g.}][]{perry:1975a,perry:1986a,davidson:2009a}
\begin{eqnarray}
 \label{eq:kmin1}
 k^+_x\phi^+_{uu} = A_1,
\end{eqnarray}
where $A_1$ is a universal constant \citep{perry:1986a}. In $(\lambda_x,z)$--space, the $k_x^{-1}$ scaling region obeys wall-scaling and outer-scaling at the small- and large-wavelength ends of the spectrum, respectively. By way of visual inspection, \citet{nickels:2005a} identified a $k_x^{-1}$ region as $z^+ > 100$, $\lambda_x > 15.7z$ and $\lambda_x < 0.3\delta$ at $Re_\tau \approx 14\,000$ (triangular region `N' in Figure~\ref{fig:spintro1}\emph{a}). To date, convincing support for a $k_x^{-1}$ dependence has not been shown, mainly due to the limited range over which this region plausibly exists. Even for the highest $Re_\tau$ laboratory data, the presence of a $k_x^{-1}$ has been inconclusive \citep{morrison:2002a,rosenberg:2013a,vallikivi:2015a,baidya:2017a}, while theory suggests that this region should grow with $Re_\tau$. A $k_x^{-1}$ scaling has been reported for atmospheric surface layer (ASL) data \citep{hogstrom:2002a,calaf:2013a} and free shear flow \citep{beresh:2018a}, however, an important issue here is that those spectra are typically presented in log-log format: $\phi_{uu} \propto k_x^{-1}$ may appear to be more convincing (Appendix~\ref{sec:appA1}). To emphasize the ambiguity of a $k_x^{-1}$ scaling, the portions of the spectra satisfying Nickels' criterion are highlighted in Figure~\ref{fig:spintro2}. These portions plateau at $A_1 \approx 0.92$ \citep{nickels:2005a}, although this plateau may be caused by a transitioning from the imprint signature of the inner-spectral peak, to the broad outer-spectral peak. Furthermore, it is well-known that the outer-spectral peak has a distinct energy footprint down to the wall, as a direct consequence of its strong linear coherence with the near-wall turbulence \citep{baars:2017a2,marusic:2017a}. These spectral features---which could enhance $\phi_{uu} \propto k_x^{-1}$ while not being truly representative of (\ref{eq:kmin1})---are most clearly observed in a spectrogram and induce a saddle-type topology (Figure~\ref{fig:spintro1}\emph{b}). \citet{chandran:2017a} examined experimentally acquired streamwise--spanwise 2D spectra of $u$ at $Re_\tau \sim \mathcal{O}(10^4)$ for a $k_x^{-1}$, but has also faced challenges in identifying an unobstructed spectral view. It is worth noting that numerical studies \citep[\emph{e.g.}][]{jimenez:2008a,lee:2015a,hwang:2015a} indicate a promising $k^{-1}$ structure in, for instance, spectra of the spanwise velocity component in channel flow, even at $Re_\tau \sim \mathcal{O}(10^3)$.

If a $k_x^{-1}$ scaling region does exist in $\phi_{uu}$, it presumably extends beyond Nickel's identified outer limit of $\lambda_x = 0.3\delta$. Any large-scale turbulence features that would scale different than $\phi_{uu} \propto k_x^{-1}$ are thought to obscure a $k_x^{-1}$ scaling at the large-scale end of the spectrum. Those obscuring features may relate to the presence of detached eddies \citep{marusic:1995a,jimenez:2012a}, the processes of spatial alignment \citep{adrian:2000a} and spectral aliasing \citep{davidson:2006ab,davidson:2008a}, all of which accumulate energy in the broad outer-spectral peak. In fact, \citet{baars:2017a2} determined that a geometrically self-similar wall-attached structure of turbulence may be ingrained within a much larger $(\lambda_x,z)$--region than the one identified by \citet{nickels:2005a}---that new region is identified region `B' in Figure~\ref{fig:spintro1}(\emph{a}).

\subsection{The spectral view of Perry et al. for streamwise velocity spectra}\label{sec:revperry}
\citet{townsend:1976bk} hypothesized that: \emph{``The velocity fields of the main eddies, regarded as persistent, organized flow patterns, extend to the wall and, in a sense, they are attached to the wall"}. For predictive purposes, Townsend further assumed a self-similar nature of the main eddies. The AEH of Townsend thus commonly implies a hierarchy of geometrically self-similar eddying motions that are inertially dominated (inviscid), attached to the wall and scalable with their distance to the wall \citep{marusic:2019a}.
\begin{figure} 
\vspace{10pt}
\centering
\includegraphics[width = 0.999\textwidth]{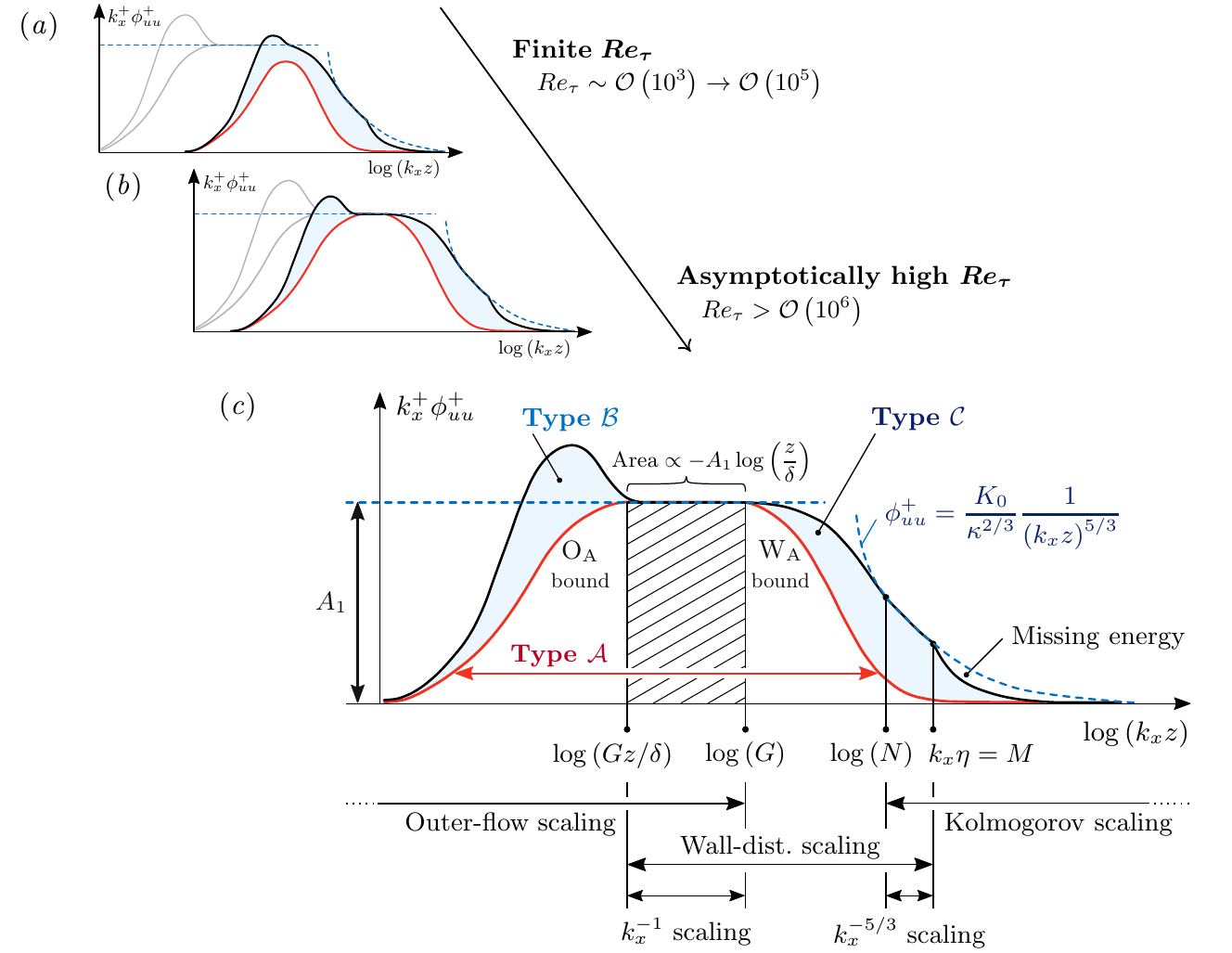}
   \caption{Hypothetical structure of $k_x\phi_{uu}$ within the logarithmic region of TBLs, following \citet{perry:1977a} and \citet{perry:1986a} and reproduced from \citet{marusic:1995a}. Sketches for finite $Re_\tau$ are hypothesized in sub-figures (\emph{a}) and (\emph{b}), while an asymptotically high $Re_\tau$ is envisioned in (\emph{c}).}
   \label{fig:spperry}
\end{figure}
In the log-region, \citet{perry:1982a} showed that an attached-eddy model predicts a $k_x^{-1}$ in $\phi_{uu}$, whereas various other theoretical arguments and interpretations lead to the same result \citep[\emph{e.g.}][]{tchen:1953a,nikora:1999a,hunt:2001a,davidson:2009a,katul:2012a}. Following the classical scaling, viscosity can be neglected above a fixed, inner-scaled wall-normal location, say $z^+ = \mathcal{O}(100)$ \citep[][among others]{perry:1982a,nagib:2007a,zagarola:1998a}. A $k_x^{-1}$ in $\phi_{uu}$ at such a lower limit of the logarithmic region was envisioned by Perry and coworkers, provided that contributions to the spectrum that do not follow a spectral overlap scaling are absent or do not overlap with a $k_x^{-1}$ range (\emph{e.g.} no imprints from the inner-spectral peak, no contributions from the scales that result in Kolmogorov scaling, and no masking by large-scale contributions that obey a $\delta$ scaling in spectral space). It must be noted here that various studies have questioned the classical scaling and have arrived at meso layer and mixed scaling concepts \citep[see][]{long:1981a,wosnik:2000a,klewicki:2009a}. However, an unresolved issue in all scaling studies for $\overline{u^2}(z)$ is that unaltered profiles are considered, which thus cluster the energy contributions from all turbulent, spectral scales (while it is agreed upon that different parts of the turbulence spectrum scale differently). Hence, a meso layer scaling of $z^+ \propto Re_\tau^{0.5}$ (the minimum wall-normal location at which viscosity acts) may not reflect the lower limit of the inviscid, logarithmic-region component of turbulence. And so, we here hypothesize following Perry and coworkers, that if one considers the scaling of sub-contributions to the streamwise velocity spectra separately, different wall-normal scalings can emerge depending on which turbulent scales are included. We highlight that a classical scaling for the lower limit of the logarithmic region in the context of inviscid, self-similar and wall-attached motions of Perry and coworkers (reviewed here in \S\,2.2) should not be discarded. Generally, the relatively slow progress on the issue of spectral scalings in TBL flow is influenced by the still limited range of scales of the energy cascade, even at moderate $Re_\tau$, thus making it complicated to observe extended scaling ranges along both wavenumber and wall-normal dimensions. For instance, with a smallest eddy on the order of $100\nu/U_\tau$, and a discrete hierarchy of 10 geometrically self-similar scales, it was shown by \citet{perry:1982a} that at least $Re_\tau \sim \mathcal{O}(10^5)$ is required for a $k_x^{-1}$ extending over one order-of-magnitude in $k_x$.

\citet{perry:1977a} and \citet{perry:1986a} contemplated that $\phi_{uu}$ encompasses three distinct contributions; a vision further refined later on \citep{marusic:1995a}. A conceptual sketch of these contributions is presented in Figure~\ref{fig:spperry}(\emph{c}) and are referred to as type $\mathcal{A}$, $\mathcal{B}$ and $\mathcal{C}$ energy contributions. Central in this view is the $\mathcal{A}$ energy (only \emph{this} energy fraction is captured by the attached-eddy model and would dominate the spectrum at ultra-high $Re_\tau$). At the high wavenumber-end, the $\mathcal{A}$ contribution follows wall-scaling with boundary W$_{\rm A}$, starting at $k_x z = G$ (note that one fixed and inner-scaled location $z$ is assumed throughout Figure~\ref{fig:spperry}). At the low wavenumber-end, $\mathcal{A}$ is bounded by O$_{\rm A}$, adhering to outer-scaling. A $k_x^{-1}$ region occurs where there is an overlap of the outer-scaling and wall-scaling. Note that both boundaries W$_{\rm A}$ and O$_{\rm A}$ are unknown and undefined, \emph{e.g.} no physical arguments exist for formulating these, nor is there empirical evidence yet for these boundaries. Logically, the spectrum also includes energy from scales that are unresolved by the attached-eddy model and that energy thus scales differently than that of $\mathcal{A}$. At low wavenumbers, $\mathcal{B}$ energy was envisioned to come from detached motions. Only through the availability of recent experimental data at high $Re_\tau$ \citep{mathis:2009a}, it appears that this contribution is Reynolds number dependent and that it induces an imprint on the near-wall region. \citet{adrian:2000a} suggested that merging of self-similar LSMs may be one of the mechanisms generating VLSMs and superstructures. But, in a recent line of work, several studies---using linearized Navier-Stokes equations---have shown that long streaky motions can be amplified at all length scales \citep[][among others]{delalamo:2006a,hwang:2010a,hwang:2010b}, and this is consistent with the recent findings that the large-scale outer structures obey a self-similar character and involve self-sustaining mechanisms \citep{hwang:2015a,degiovanetti:2017a}. \citet{vassilicos:2015a} performed a modelling attempt of this low wavenumber-end of the spectrum at high $Re_\tau$, by including a $k_x^{-m}$ range. In the current context this can be interpreted as an attempt to model the $\mathcal{B}$ contribution. In addition, \citet{srinath:2018a} related this modelling effort to the wide variation of streamwise lengths of wall-attached streamwise velocity structures. Finally, at the high wavenumber-end in Figure~\ref{fig:spperry}(\emph{c}), region $\mathcal{C}$ comprises Kolmogorov scales and small-scale detached eddies. This portion of the spectrum obeys $k_x^{-5/3}$ and is subject to the viscous roll-off beyond an $\eta$-based scale (boundary with constant $M$).

Perhaps the most important feature in Figure~\ref{fig:spperry}(\emph{c}) is the complete spectral separation of $\mathcal{B}$ and $\mathcal{C}$ components at ultra-high $Re_\tau$, which has never been observed. The significant spectral overlap envisioned at low Reynolds number conditions, with a small range of energetic length scales, is schematically shown in Figures~\ref{fig:spperry}(\emph{a,b}). In essence, at low $Re_\tau$, the range of scales has not yet matured to a range where complete $\mathcal{B}$/$\mathcal{C}$ spectral-separation may occur (if at all existent). In light of the above, this article attempts to re-appraise the spectral view of Figure~\ref{fig:spperry} by way of utilizing data over a range of $Re_\tau$, in combination with a data-driven spectral decomposition technique.


\subsection{Present contribution and outline}\label{sec:outline}
To summarize the above, a presence of geometrically self-similar, wall-attached motions, as hypothesized by \citet{townsend:1976bk}, would imply the existence of $\phi_{uu} \propto k_x^{-1}$ at high enough $Re_\tau$. This attached-eddy model behaviour only governs $\mathcal{A}$ in the spectral view hypothesized by Perry and co-workers. Empirical validation of a three-component structure in the velocity spectra $\phi_{uu}$ is non-existent, and boundaries O$_{\rm A}$ and W$_{\rm A}$, as well as constants like $G$, $N$ and $M$, are unknown (Figure~\ref{fig:spperry}). Moreover, a convincing $\phi_{uu} \propto k_x^{-1}$ scaling has remained elusive as it has not been observed in high-fidelity data. In order to gain further insight into the structure of $\phi_{uu}$, there is a need to examine the possible existence of a three-component spectral structure. This article presents a first attempt towards this, using a data-driven spectral decomposition method. Next, in \S\,\ref{sec:data}, a description of the multi-point synchronized experimental data is provided. Via linear systems theory (\S\,\ref{sec:coherence}), data-driven spectral filters are derived (\S\S\,\ref{sec:nearwall}--\ref{sec:logregion}) and allow for the spectral decomposition of $\phi_{uu}$ (\S\,\ref{sec:trdecom}). It also provides further insight into the required scale separation for observance of $\phi_{uu} \propto k_x^{-1}$ (\S\,\ref{sec:kmin1}). Subsequently, spectra at a range of $Re_\tau$ are decomposed and interpreted according to the spectral structure of Perry and co-workers (\S\,\ref{sec:Retrend}). In Part~2 \citep{baars:part2}, the integrated energy in the $\phi_{uu}$ spectra, being the streamwise turbulence intensity $\overline{u^2}$, is reassessed in the context of the spectral decomposition presented in this paper.

\section{Turbulent boundary layer data}\label{sec:data}
\begin{figure} 
\vspace{10pt}
\centering
\includegraphics[width = 0.999\textwidth]{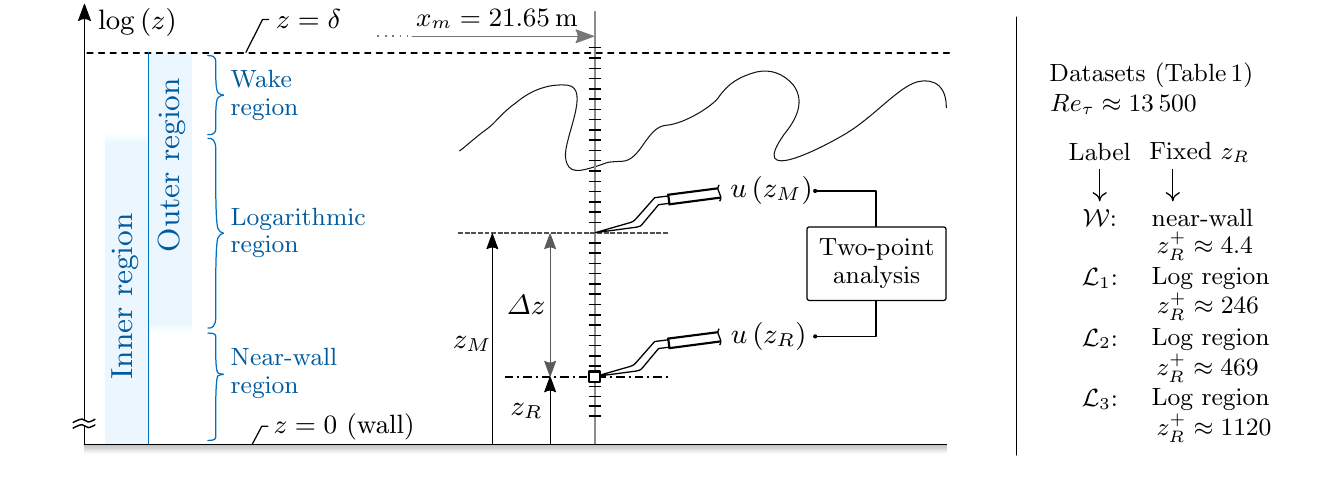}
   \caption{Experimental arrangement of the two-probe synchronized hot-wire measurements in Melbourne's boundary facility. TBL conditions and probe-positions are listed in Table~\ref{tab:expcond}.}
   \label{fig:expsetup}
\end{figure}
\subsection{Two-point measurements at $Re_\tau \approx 13\,500$}\label{sec:twopoint}
Two-point measurement data are the foundation for the spectral decomposition-methodology used. Data were acquired in Melbourne's boundary layer facility \citep{nickels:2005a,baars:2016aa} under nominal ZPG conditions. Pressure coefficient, $C_p$, was constant to within $\pm0.87$\,\% \citep{marusic:2015a}, and free-stream turbulence intensities were less than 0.05\,\% of the free-stream at $x = 0$\,m \citep{kulandaivelu:2011phd}.

\begin{figure} 
\vspace{10pt}
\centering
\includegraphics[width = 0.999\textwidth]{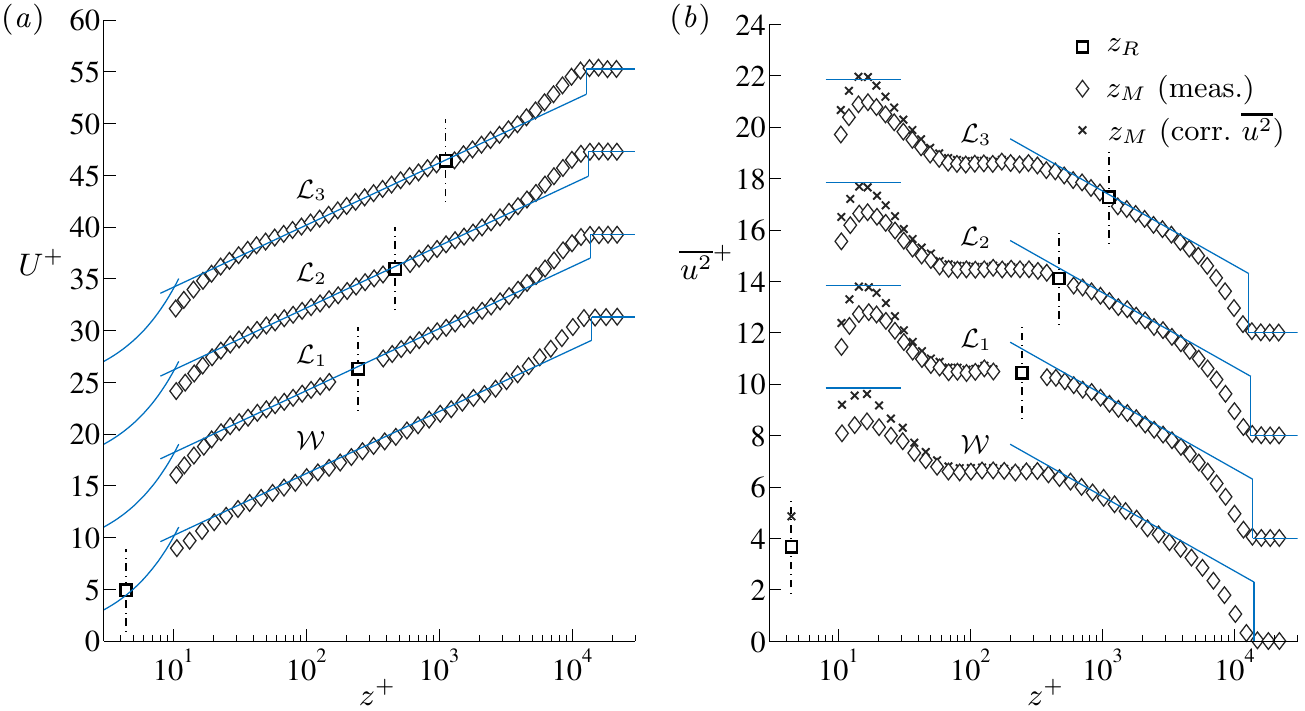}
   \caption{Boundary layer profiles of the streamwise (\emph{a}) mean velocity and (\emph{b}) turbulence intensity of the four two-point hot-wire datasets listed in Table~\ref{tab:expcond}.}
   \label{fig:expprof}
\end{figure}

Four datasets were acquired using a hot-wire arrangement shown in Figure~\ref{fig:expsetup}. For each dataset, one hot-wire probe is positioned at a fixed location (reference $z = z_R$), while the other probe is traversed to map out a range of wall-normal positions, $z_M$ (probes were located at the same $(x,y)$--position). One dataset (denoted by $\mathcal{W}$) employs a near-wall reference at $z^+_R \approx 4.4$, while the reference probe was situated within the logarithmic region for the other three datasets ($\mathcal{L}_i$, $i = 1...3$, with reference positions $z^+_R \approx 246$, $469$ and $1\,120$, respectively). These four reference locations are from now on denoted as $z_{\mathcal{W}}$ and $z_{\mathcal{L}}$ for datasets $\mathcal{W}$ and $\mathcal{L}_{1-3}$, respectively. Nominally, all datasets correspond to the same Reynolds number condition ($Re_\tau \approx 13\,500$). Boundary layer parameters are summarized in Table~\ref{tab:expcond}. The friction velocity was inferred from a direct survey of the wall-shear stress using a floating element drag balance, situated at the streamwise measurement location \citep{baars:2016aa}, and a modified Coles law of the wake fit was used to determine $\delta$ \citep{jones:2001a}. For reference, mean velocity profiles (Figure~\ref{fig:expprof}\emph{a}) were fitted to a composite profile with log-law constants of $\kappa = 0.384$ and $A = 4.17$ \citep{chauhan:2009a}. This procedure resulted in values for $U_\tau$ that deviated less than 2\,\% from the friction data-based values. 
\begin{table}
  \begin{center}
  \begin{minipage}{\textwidth}  
  \vspace*{-6pt}
  \begin{tabular}{@{}llcccccccc@{}}
  \multicolumn{5}{l}{\textbf{Turbulent boundary layer:}}\\[2pt]
  ~ & Datasets & Label & \multirow{2}{*}{$Re_\tau \equiv \dfrac{\delta U_\tau}{\nu}$} & \multirow{2}{*}{$Re_\theta \equiv \dfrac{\theta U_\infty}{\nu}$} & $x_m$ & $U_{\infty}$ & $\delta$ & $U_\tau$ & $\nu/U_\tau$ \\
  ~ & ~ & ~ & ~ & ~ & (m) & (ms$^{-1}$) & (mm) & (ms$^{-1}$) & ($\upmu$m)\\[2pt]
  ~ & \,near-wall ref. & $\mathcal{W}$ & 14\,100 & 36\,100 & 21.65 & 19.95 & 338 & 0.638 & 24.0\\[2pt]
  ~ & \ldelim\{{3}{20mm}[log-region ref.] & $\mathcal{L}_1$ & 13\,700 & 36\,000 & 21.65 & 20.37 & 331 &  0.651 & 24.1\\[1.2pt]
  ~ & ~ & $\mathcal{L}_2$ & 13\,300 & 35\,900 & 21.65 & 20.35 & 320 & 0.651 & 24.1\\[1.2pt]
  ~ & ~ & $\mathcal{L}_3$ & 12\,800 & 35\,700 & 21.65 & 20.15 & 312 & 0.645 & 24.3\\[5pt]
  \end{tabular}
  \begin{tabular}{@{}llcccccccccccc@{}}
  \multicolumn{5}{l}{\textbf{Hot-wire anemometry:}}\\[2pt]
  ~ & ~ & ~ & \multicolumn{4}{l}{reference wire$^\#$} & ~ & \multicolumn{3}{l}{traversing wire$^\#$} & ~ & \multicolumn{2}{l}{acquisition} \\\cmidrule{4-7}\cmidrule{9-11}\cmidrule{13-14}
  ~ & Datasets & Label & $z^+_R$ & $z_R/\delta$ & $l^+$ & $l/d$ & ~ & $z^+_M$\,{[min--max]} & $l^+$ & $l/d$ & ~ & $\Delta T^+$ & $T U_{\infty}/\delta$\\[2pt]
  ~ & \,near-wall ref. & $\mathcal{W}$ & 4.4 & 0.0003 & 42 & 200 & ~ & 10.7\,--\,1.55$\delta^+$ & 22 & 200 & ~ & 1.33 & 21\,200\\[2pt]
  ~ & \ldelim\{{3}{20mm}[log-region ref.] & $\mathcal{L}_1$ & 246 & 0.0179 & 21 & 200 & ~ & 10.6\,--\,1.59$\delta^+$ & 21 & 200 & ~ & 0.60 & 20\,300\\[1.2pt]
  ~ & ~ & $\mathcal{L}_2$ & 469 & 0.0353 & 21 & 200 & ~ & 10.5\,--\,1.64$\delta^+$ & 21 & 200 & ~ & 0.61 & 22\,900\\[1.2pt]
  ~ & ~ & $\mathcal{L}_3$ & 1120 & 0.0871 & 21 & 200 & ~ & 10.4\,--\,1.68$\delta^+$ & 21 & 200 & ~ & 0.60 & 23\,200\\
  \end{tabular}
  \caption{Experimental parameters of two-point hot-wire data acquired in Melbourne's boundary layer
facility. $^\#$\,For datasets $\mathcal{L}_{1-3}$ the reference and traversing wires were interchanged for $z_M < z_R$ and $z_M > z_R$, see text for details.}
  \label{tab:expcond}
  \end{minipage}
  \end{center}
\end{table}

For dataset $\mathcal{W}$, the near-wall probe was a Dantec 55P30 type, while the traversing hot-wire consisted of a Dantec 55P15 boundary layer probe. Platinum-Wollaston wires were used with a $d = 5$\,$\upmu$m and $d = 2.5$\,$\upmu$m core wire, for the near-wall and traversing probes, respectively. Both wires were etched to expose the core over a length of $l/d = 200$ \citep{ligrani:1987a}. In viscous-scaling, the lengths of $l^+ \approx 42$ and $l^+ \approx 22$ are of an acceptable length when the spatial resolution is concerned, since this study concentrates on the absolute energy in the logarithmic region only \citep{hutchins:2009a,samie:2018a}. The traversing hot-wire was traversed to 40 logarithmically-spaced positions in the range $10.7 \apprle z^+_M \apprle 1.55\delta^+$. For datasets $\mathcal{L}_{1-3}$, both probes consisted of Dantec 55P15 type holders with $d = 2.5$\,$\upmu$m wires (also $l/d = 200$, resulting in $l^+ \approx 21$). Each probe could be moved independently in $z$ using two separate traverse systems: a tunnel-specific traverse with a wall-implemented traverse situated underneath. Irrespective of $z_R$, a mapping of 50 logarithmically-spaced positions from $z^+_M \approx 10.5$ up to $z^+_M \approx 1.6\delta^+$ was carried out. In practice, positions below $z_R$ were mapped out with the probe connected to the bottom traverse (with the other probe at $z_R$). Subsequently, the probes switched positions, and the wall-normal range above $z_R$ was mapped out with the top traverse (with the bottom traverse-probe at $z_R$). Since the geometry of the probe holders dictated a minimum separation distance of $\Delta z^+_{\rm min} = \vert z_M - z_R \vert \approx 2.2$\,mm, 5 points ($\mathcal{L}_1$), 2 points ($\mathcal{L}_2$) and 1 point ($\mathcal{L}_3$) were skipped in the 50-point profiles. 

All hot-wire probes were operated in constant temperature mode, with an overheat ratio of 1.8, using in-house built anemometers. For each dataset, wires were sampled simultaneously at a rate of $\Delta T^+ \equiv U^2_\tau/\nu/f_s$, where $f_s$ is the sampling frequency \citep[acquisition rates, $\Delta T^+$, were around unity or less, see][]{hutchins:2009a}. To prevent aliasing, the signals were passed through 4$^{\rm th}$ order Butterworth filters---with a spectral cut-off set at $f_s/2$---prior to A/D conversion using a 16-bit Data Translation DT9836 module. Relatively long signals were acquired with lengths of $TU_{\infty}/\delta > 20 \times 10^3$, allowing for converged spectral statistics at the largest energetic wavelengths. Both hot-wire probes were calibrated, with a correction method for hot-wire voltage drift \citep{talluru:2014ba}. 
Boundary layer profiles of $U$ and $\overline{u^2}$ are plotted in Figures~\ref{fig:expprof}(\emph{a}) and~\ref{fig:expprof}(\emph{b}), respectively. All mean velocity profiles are compared with $U^+ = 1/\kappa\ln(z^+) + A$ in the logarithmic region (thin blue lines), with $\kappa = 0.384$ and $A = 4.17$ and show minimum effect from the intrusive reference probe. Likewise, the agreement of the $\overline{u^2}$ profiles can be gleaned from their comparison to $\overline{u^2}^+ = B_1 - A_1\ln(z/\delta)$ in the logarithmic region (thin blue lines), with $A_1 = 1.26$ and $B_1 = 2.30$ \citep{marusic:2013a}; profiles $\mathcal{L}_{1-3}$ comprise a slightly lower energy in the outer-region than the $\mathcal{W}$ data, which is ascribed to the minor variation in $Re_\tau$. Since the near-wall streamwise turbulence intensity is attenuated due to the hot-wire's spatial resolution, corrected profiles via the method of \citet{smits:2011a} are also shown. Peak-values at $z^+ \approx 15$ agree with the expected behaviour following $\overline{u^2}_{\rm max} = 0.63 \ln(Re_\tau) + 3.80$ \citep[][]{lee:2015a,marusic:2017a,samie:2018a}.

\subsection{Additional data for investigating Reynolds number dependence}\label{sec:Redata}
A number of other TBL datasets are employed for investigating Reynolds number trends, ranging from DNS up to ASL data. For all datasets, $\delta$, $U_\tau$ and $Re_\tau$ were recomputed with the modified Coles law of the wake fit. DNS data of a ZPG TBL by \citet{sillero:2013a} correspond to $Re_\tau \approx 1\,992$. Streamwise/wall-normal planes of data, spanning the entire TBL in $z$ and extending $\sim 11.9\delta$ in $x$, are used. This streamwise extension is chosen to include large-scale motions, while still maintaining an acceptable Reynolds number increase from $Re_\theta \equiv \theta U_{\infty}/\nu \approx 5\,110$ to $6\,010$, where $\theta$ is the momentum thickness ($Re_\tau \approx 1992$ at the streamwise centre). Around this Reynolds number condition, the spanwise resolution of the DNS is $\Delta y^+ \approx 3.72$.

Five singe-point hot-wire datasets of $u$ are also included throughout this work (all acquired in Melbourne's TBL facility). Data at $Re_\tau \approx 2\,800$, $3\,900$ and $7\,300$ were taken using a regular hot-wire set-up \citep{hutchins:2009a}. Data at $Re_\tau \approx 13\,000$ and $19\,300$ are taken from \citet{samie:2018a}, where Princeton's NSTAP probes were used for making fully-resolved measurements. Our highest Reynolds number data at $Re_\tau \approx 1.4 \times 10^6$ encompass the $u$ fluctuation in the ASL under near-neutrally buoyant conditions at the SLTEST on the great salt lakes of Utah. These data were synchronously acquired using a wall-normal array of five sonic anemometers and one purpose-built wall-shear stress sensor situated under the array \citep{heuer:2005a,marusic:2007a}.

\section{Data-driven spectral filters for an energy decomposition}\label{sec:filters}
To generate data-driven spectral filters from the two-point data, spectral coherence analyses (\S\,\ref{sec:coherence}) are applied to the dataset with the near-wall reference signal (\S\,\ref{sec:nearwall}), as well as the datasets with the logarithmic-region reference (\S\,\ref{sec:logregion}).

\subsection{Methodology}\label{sec:coherence}
\subsubsection{Linear coherence spectrum: a data-driven filter}\label{sec:lcs}
A stochastic coupling between two time series of $u$ can be inferred from correlation analyses. Instead of computing the correlation in physical space, as the two-point cross-correlation coefficient $\rho(\tau)$, where $\tau$ is the relative time shift between the two signals, a spectral approach is pursued for its relevance to energy spectra. A diagnostic for the scale-by-scale coupling is the linear coherence spectrum \citep[LCS,][]{bendat:1986bk}. For two synchronously acquired time series $u(z)$ and $u(z_R)$, the LCS is formulated as
\begin{eqnarray}
 \label{eq:LCS}
 \gamma^2_l\left(z,z_R;\lambda_x\right) \equiv \frac{\left\vert \left\langle \widehat{u}\left(z;\lambda_x\right) \overline{\widehat{u}\left(z_R;\lambda_x\right)} \right\rangle \right\vert^2}{\left\langle \left\vert \widehat{u}\left(z;\lambda_x\right) \right\vert^2\right\rangle \left\langle \left\vert \widehat{u}\left(z_R;\lambda_x\right) \right\vert^2\right\rangle} = \frac{\left\vert\phi'_{uu}\left(z,z_R;\lambda_x\right)\right\vert^2}{\phi_{uu}\left(z;\lambda_x\right) \phi_{uu}\left(z_R;\lambda_x\right)},
\end{eqnarray}
where $\widehat{u}\left(z;\lambda_x\right)$ is the temporal Fourier transform of $u(z)$. Although $\lambda_x$ is here used as the scale-argument, the expression is evaluated in frequency space. The overbar in (\ref{eq:LCS}) indicates the complex conjugate, $\langle \rangle$ denotes ensemble averaging and $\vert \vert$ designates the modulus. Since $\gamma^2_l$ equals the cross-spectrum magnitude-squared, normalized by the energy spectra of $u(z)$ and $u(z_R)$, a normalized coherence is obtained: $0 \leq \gamma^2_l \leq 1$. Interpretation-wise, $\gamma^2_l$ reflects the square of a scale-specific correlation coefficient and represents the fraction of common variance shared by $u(z)$ and $u(z_R)$, per scale. Although only the magnitude of the complex-valued cross-spectrum, $\phi'_{uu}(z,z_R;\lambda_x) \in \mathbb{C}$, is considered in the nominator of (\ref{eq:LCS}), the LCS implicitly embodies the phase consistency (and co-existing amplitude variations) across ensembles of $u(z)$ and $u(z_R)$. That is, if each ensemble used to construct the cross-spectrum contains a radically different (non-consistent) phase shift for a certain scale, that scale is not stochastically correlated and $\gamma^2_l \rightarrow 0$ when converged. 
\begin{figure} 
\vspace{10pt}
\centering
\includegraphics[width = 0.999\textwidth]{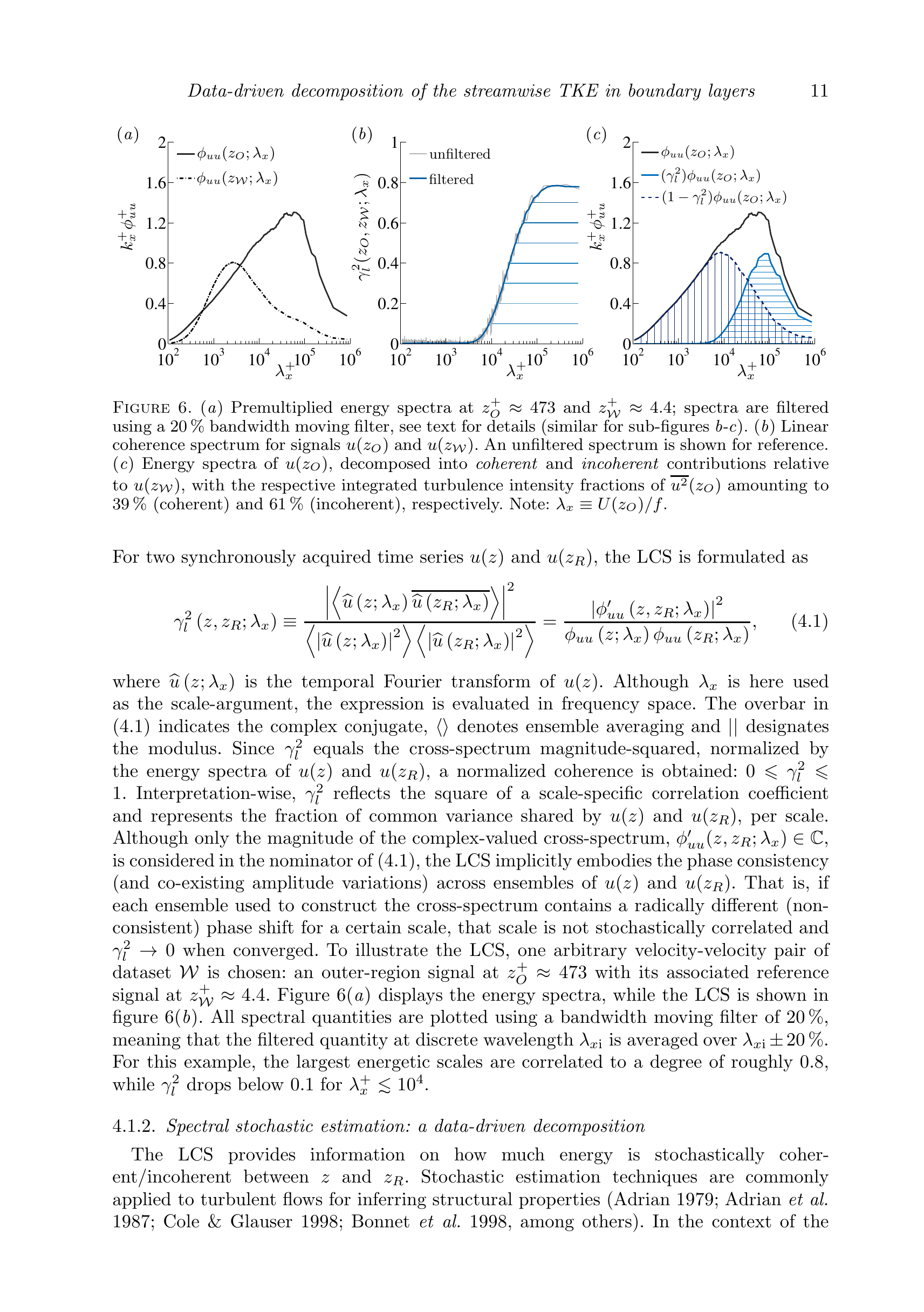}
   \caption{(\emph{a}) Premultiplied energy spectra at $z^+_O \approx 473$ and $z^+_{\mathcal{W}} \approx 4.4$; spectra are filtered using a 20\,\% bandwidth moving filter, see text for details (similar for sub-figures \emph{b-c}). (\emph{b}) Linear coherence spectrum for signals $u(z_O)$ and $u(z_{\mathcal{W}})$. An unfiltered spectrum is shown for reference. (\emph{c}) Energy spectra of $u(z_O)$, decomposed into \emph{coherent} and \emph{incoherent} contributions relative to $u(z_{\mathcal{W}})$, with the respective integrated turbulence intensity fractions of $\overline{u^2}(z_O)$ amounting to 39\,\% (coherent) and 61\,\% (incoherent), respectively. Note: $\lambda_x \equiv U(z_O)/f$.}
   \label{fig:cohdecEX}
\end{figure}
To illustrate the LCS, one arbitrary velocity-velocity pair of dataset $\mathcal{W}$ is chosen: an outer-region signal at $z^+_O \approx 473$ with its associated reference signal at $z^+_{\mathcal{W}} \approx 4.4$. Figure~\ref{fig:cohdecEX}(\emph{a}) displays the energy spectra, while the LCS is shown in figure~\ref{fig:cohdecEX}(\emph{b}). All spectral quantities are plotted using a bandwidth moving filter of 20\,\%, meaning that the filtered quantity at discrete wavelength $\lambda_{x\rm{i}}$ is averaged over $\lambda_{x\rm{i}}\pm 20$\,\%. For this example, the largest energetic scales are correlated to a degree of roughly 0.8, while $\gamma^2_l$ drops below 0.1 for $\lambda^+_x \apprle 10^4$.

\subsubsection{Spectral stochastic estimation: a data-driven decomposition}\label{sec:lse}
The LCS provides information on how much energy is stochastically coherent/incoherent between $z$ and $z_R$. Stochastic estimation techniques are commonly applied to turbulent flows for inferring structural properties \citep[][among others]{adrian:1979a,adrian:1987a,cole:1998a,bonnet:1998a}. In the context of the LCS, and considering $u(z_R)$ and $u(z)$ as an input and output of a linear time-invariant system, respectively, a spectral linear stochastic estimate (LSE) of the output is \citep{bendat:1986bk,tinney:2006a}
\begin{eqnarray}
 \label{eq:SE}
 \widehat{u}^E\left(z;\lambda_x\right) = H_l\left(z,z_R;\lambda_x\right)\widehat{u}\left(z_R;\lambda_x\right).
\end{eqnarray}
The linear transfer kernel $H_l$ is computed from an ensemble of data via
\begin{eqnarray}
 \label{eq:HL}
 H_l\left(z,z_R;\lambda_x\right) = \frac{\phi'_{uu}\left(z,z_R;\lambda_x\right)}{\phi_{uu}\left(z_R;\lambda_x\right)} = \left\vert H_l \right\vert e^{j\varphi} \in \mathbb{C},
\end{eqnarray}
comprising a wavelength-dependent linear gain, $\vert H_l \vert$, and phase $\varphi$. As is evident, the spectral LSE approach via (\ref{eq:SE}) transforms a frequency-domain unconditional input into an estimate of the frequency-domain conditional output through one multiplicative step. In essence, spectral LSE is an efficient implementation of the multi-time LSE scheme \citep[for details:][]{ewing:1999c,tinney:2006a}. Note that a linear stochastic estimate of the output time series, $\widehat{u}(z)$ can simply be obtained from the inverse Fourier transform of (\ref{eq:SE})---see also \citet{baars:2016ab}. By combining (\ref{eq:LCS}) and (\ref{eq:HL}), a stochastic estimate of the output's energy spectrum can be expressed as
\begin{eqnarray}
 \label{eq:coh}
 \phi^E_{uu}\left(z;\lambda_x\right) = \left\vert H_l\left(z,z_R;\lambda_x\right) \right\vert^2 \phi_{uu}\left(z_R;\lambda_x\right) = \gamma^2_l\left(z,z_R;\lambda_x\right) \phi_{uu}\left(z;\lambda_x\right).
\end{eqnarray}
This implies that the amount of energy at $z$ that can be reconstructed via an LSE procedure, from an input at $z_R$, is equal to the measured spectrum at $z$, multiplied by $\gamma^2_l$. And thus, it is said that the estimated spectrum comprises the coherent portion of the actual spectrum (relative to a reference location from which the estimate is performed). Consequently, the LCS can be used as a wavelength-dependent filter for decomposing $\phi_{uu}(z;\lambda_x)$ into stochastically \emph{coherent} and \emph{incoherent} portions, relative to $z_R$, following
\begin{eqnarray}
 \label{eq:dec}
 \phi_{uu}\left(z;\lambda_x\right) = \underbracket{\left(\gamma^2_l\right)\phi_{uu}\left(z;\lambda_x\right)}_{\text{\scalebox{1.15}{coherent: $\phi^E_{uu}$}}} + \underbracket{\left(1 - \gamma^2_l\right)\phi_{uu}\left(z;\lambda_x\right)}_{\text{\scalebox{1.15}{incoherent}}}.
\end{eqnarray}
Figure~\ref{fig:cohdecEX}(\emph{c}) illustrates this decomposition with visuals of the coherent and incoherent portions of $\phi_{uu}(z_O;\lambda_x)$, comprising $\sim 39$\,\% and $\sim 61$\,\% of the total turbulence intensity, $\overline{u^2}(z_O)$, respectively. Because the input/output in (\ref{eq:SE})--(\ref{eq:dec}) is reversible, $\phi_{uu}(z_R;\lambda_x)$ can also be decomposed into coherent and incoherent portions. 

It is noted that the method above is linear only. \citet{guezennec:1989a} showed that the inclusion of quadratic terms in stochastic estimates of the velocity fields of a TBL flow---from a velocity input---negligibly improves the estimate. Furthermore, \citet{naguib:2001a} did show that the inclusion of higher-order terms does improve the estimate of the velocity field from surface pressure as input (via a time-domain quadratic stochastic estimation scheme), but, this was linked to an inherent nonlinearity caused by the turbulent--turbulent pressure source term. Only velocity-velocity coupling is considered in this work, hence a linear technique is sufficient. 

\subsection{Coherence relative to a near-wall reference}\label{sec:nearwall}
\subsubsection{Coherence spectrogram and physical interpretation}\label{sec:wallcoh}
\begin{figure} 
\vspace{10pt}
\centering
\includegraphics[width = 0.999\textwidth]{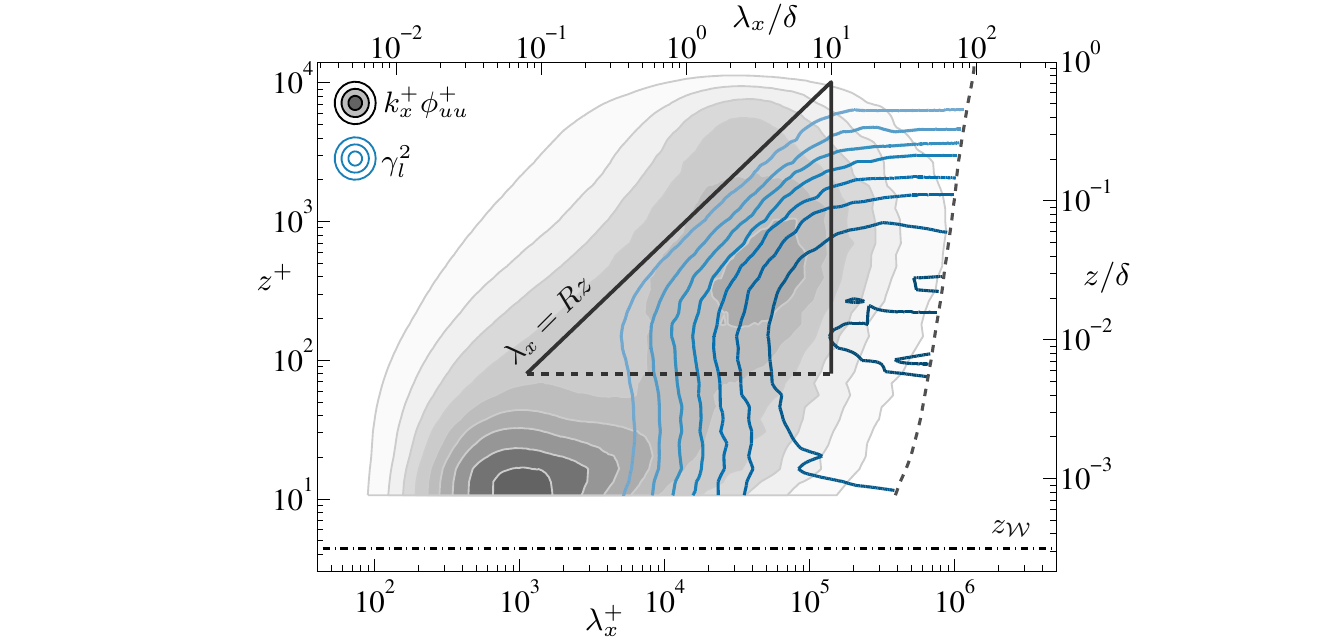}
   \caption{Dataset $\mathcal{W}$. Coherence spectrogram $\gamma^2_l(z,z_{\mathcal{W}};\lambda_x)$, relative to the near-wall reference location $z^+_{\mathcal{W}} \approx 4.4$ (iso-contours 0.1:0.1:0.9), superposed on its associated premultiplied energy spectrogram $k^+_x\phi^+_{uu}$ (filled iso-contours 0.2:0.2:1.8), reproduced from Figure~\ref{fig:spintro1}. \citet{baars:2017a2} contemplated a self-similar behaviour in the triangular region (here drawn with $R \approx 14$).} 
   \label{fig:LCSdataW}
\end{figure}
Using (\ref{eq:LCS}), coherence spectra can be computed from the $\mathcal{W}$ dataset for all $z$ within the TBL. A coherence spectrogram, formed by presenting all individual coherence spectra as iso-contours of $\gamma^2_l$, is presented in $(\lambda_x,z)$--space and superposed on the $k_x^+\phi^+_{uu}$ energy spectrogram in Figure~\ref{fig:LCSdataW}. A horizontal cut through the $\gamma^2_l$ contour at $z^+_O \approx 473$ results in Figure~\ref{fig:cohdecEX}(\emph{b}). Iso-contours of $\gamma^2_l$ increase in value, with increasing $\lambda_x$, and follow lines of constant $\lambda_x/z$ within the logarithmic region. Only a portion of the energy spectrogram below non-zero contours of $\gamma^2_l$ is stochastically coherent with $z_{\mathcal{W}}$. A dashed line at the large-wavelength end indicates the lowest frequency resolved.
\begin{figure} 
\vspace{10pt}
\centering
\includegraphics[width = 0.999\textwidth]{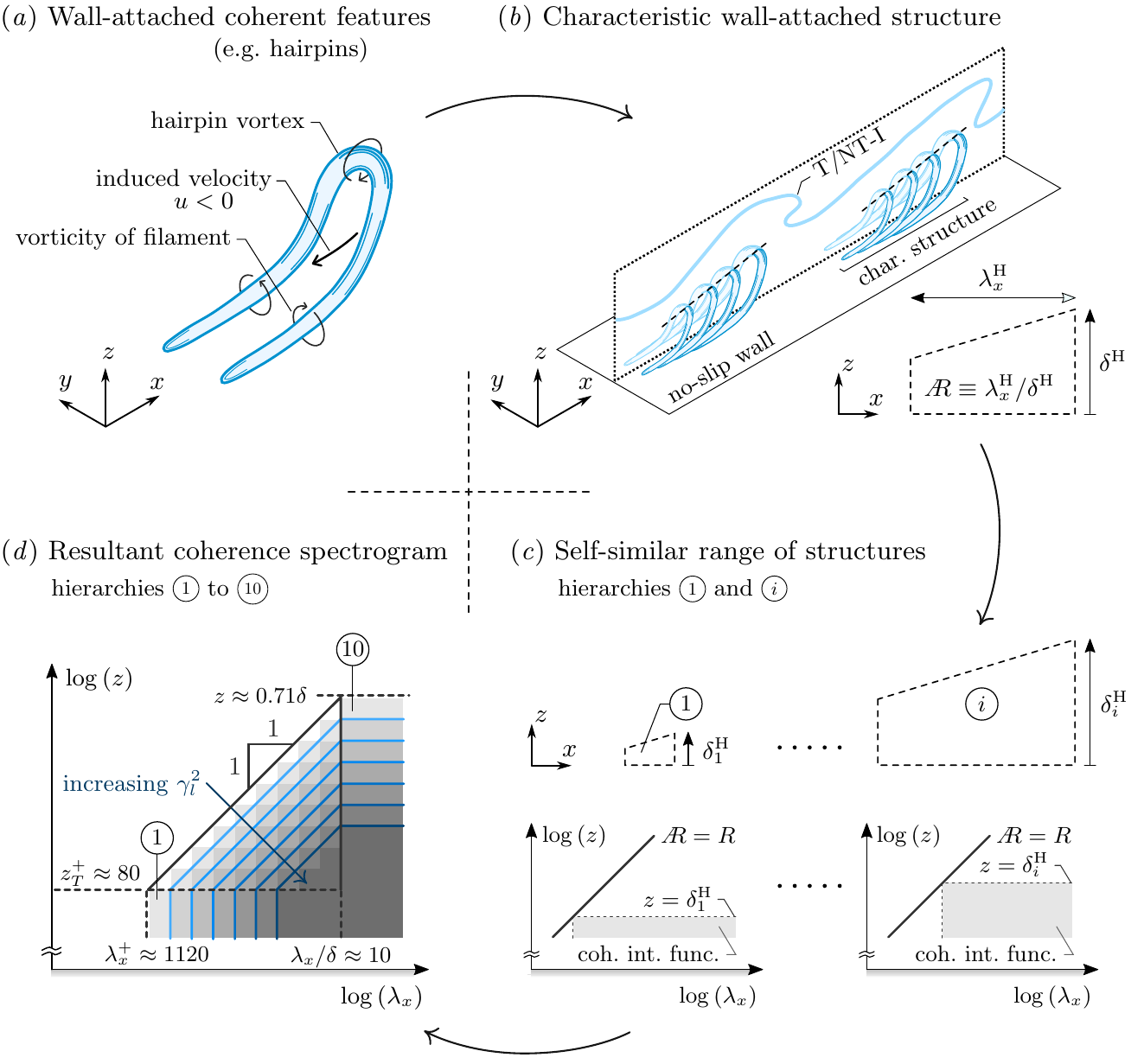}
   \caption{(\emph{a}-\emph{d}) A visual roadmap conceptualizing a physical underpinning for a coherence spectrogram with a near-wall reference location. Precise forms of the structures in (\emph{a}) and (\emph{b}), drawn as a hairpin vortex and packets of hairpins, are of secondary importance in the context of a hierarchical distribution of \emph{some} wall-attached eddies, and solely serves for visual purposes.}
   \label{fig:roadmapW}
\end{figure}

A physical interpretation for the $\gamma^2_l$ trend is considered here with the aid of a roadmap in Figures~\ref{fig:roadmapW}(\emph{a}-\emph{d}). Here we assume the existence of a wall-attached structure with an embedded self-similar hierarchy of scales \citep{baars:2017a2}. Figure~\ref{fig:roadmapW}(\emph{a}) visualizes a wall-attached coherent feature as a hairpin vortex with a forward-leaning inclination angle of $\sim 45^{\circ}$ \citep{adrian:2000a}. Note that the exact types of structures do not matter in this discussion. Streamwise and wall-normal extents of such a structure are roughly equal, implying a streamwise/wall-normal aspect ratio of $\mathcal{O}(1)$. When the hairpins form packets (Figure~\ref{fig:roadmapW}\emph{b}), their streamwise extent grows more than their wall-normal region of influence \citep[and a characteristic angle reduces to $\sim 12^{\circ}$, \emph{e.g.}][]{christensen:2001a}. The wall-attached structures have a characteristic wavelength $\lambda_x^H$ and wall-normal extent $\delta_H$, so that a streamwise/wall-normal aspect ratio arises as $\AR \equiv \lambda_x^H/\delta_H$. The forward-leaning nature of the structures is neglected since a consistent phase shift is irrelevant in the context of the LCS and spectra. Townsend's self-similarity implies that each structure-hierarchy is associated with the same $\AR$ and two structures are shown in Figure~\ref{fig:roadmapW}(\emph{c}) of heights $\delta_1^H$ and $\delta_i^H$. Due to the structures' random repetitions in space (or time) \citep{woodcock:2015a}, we here assume a non-zero coherence for $\lambda_x > \lambda_x^H$: thus, $\gamma^2_l > 0$ for $z < \delta_i^H$ and $\lambda_x > \AR \delta_i^H$ in the case of hierarchy $i$ (bottom of Figure~\ref{fig:roadmapW}\emph{c}). The magnitude of $\gamma^2_l$ may vary with $\lambda_x$ and $z$, but a constant magnitude is drawn for ease with a single grey-scale. A non-uniform magnitude results in the same final structure of the reconstructed $\gamma^2_l$ spectrogram when $Re_\tau \rightarrow \infty$ (an infinite number of hierarchies). Here the magnitude of the coherence intensity function is less than unity and is attributed to an inherent three-dimensionality of wall-attached structures (only coherence in the $(x,z)$--plane is considered), stochastically inconsistent motions (\emph{e.g.} random meandering or flapping) and its continuous evolution (growth, saturation, decay). These factors contribute to the trivial cause of less-than-perfect coherence: the co-existence of detached (wall-incoherent) energy in $(\lambda_x,z)$--space. This incoherent energy does add to the total energy in the denominator of (\ref{eq:LCS}). Finally, when superposing each hierarchy of self-similar scales to reconstruct a coherence spectrogram (Figure~\ref{fig:roadmapW}\emph{d}), $\gamma^2_l$ increases in the region where individual coherence intensity functions overlap, allegedly caused by the growing ratio of attached energy, relative to detached energy (supported by the data and here visualized by the increasing grey scale of the superposed transparent hierarchy of rectangles from Figure~\ref{fig:roadmapW}\emph{c}). In a region bounded by $z = l$, a constant $\lambda_x/z$ and a constant $\lambda_x$, the $\gamma^2_l$ iso-contours thus obey a constant $\lambda_x/z$ via
\begin{eqnarray}
 \label{eq:baars}
 \gamma^2_l = C_1\ln\left(\frac{\lambda_x}{z}\right) + C_2 = C_1\ln\left(\frac{\lambda_x}{z}\frac{1}{R}\right),
\end{eqnarray}
and $\gamma^2_l$ is bounded: $\gamma^2_l \in [0,1]$. Constant $R$ is the aspect ratio at which the coherence falls to zero, \emph{e.g.} $R = \lambda_x/z\vert_{\gamma^2_l \rightarrow 0} = \exp(\text{-}C_2/C_1)$. This example with $l^+ \approx 80$ and $z \approx 0.71\delta$ for the upper bound of the triangular region \citep{baars:2017a2}, and with the assumption that each of the ten discrete hierarchies are subject to a doubling in size, yields $Re_\tau = 80 \cdot 2^{(10-1)}/0.71 \approx 58\,000$.

\begin{figure} 
\vspace{10pt}
\centering
\includegraphics[width = 0.999\textwidth]{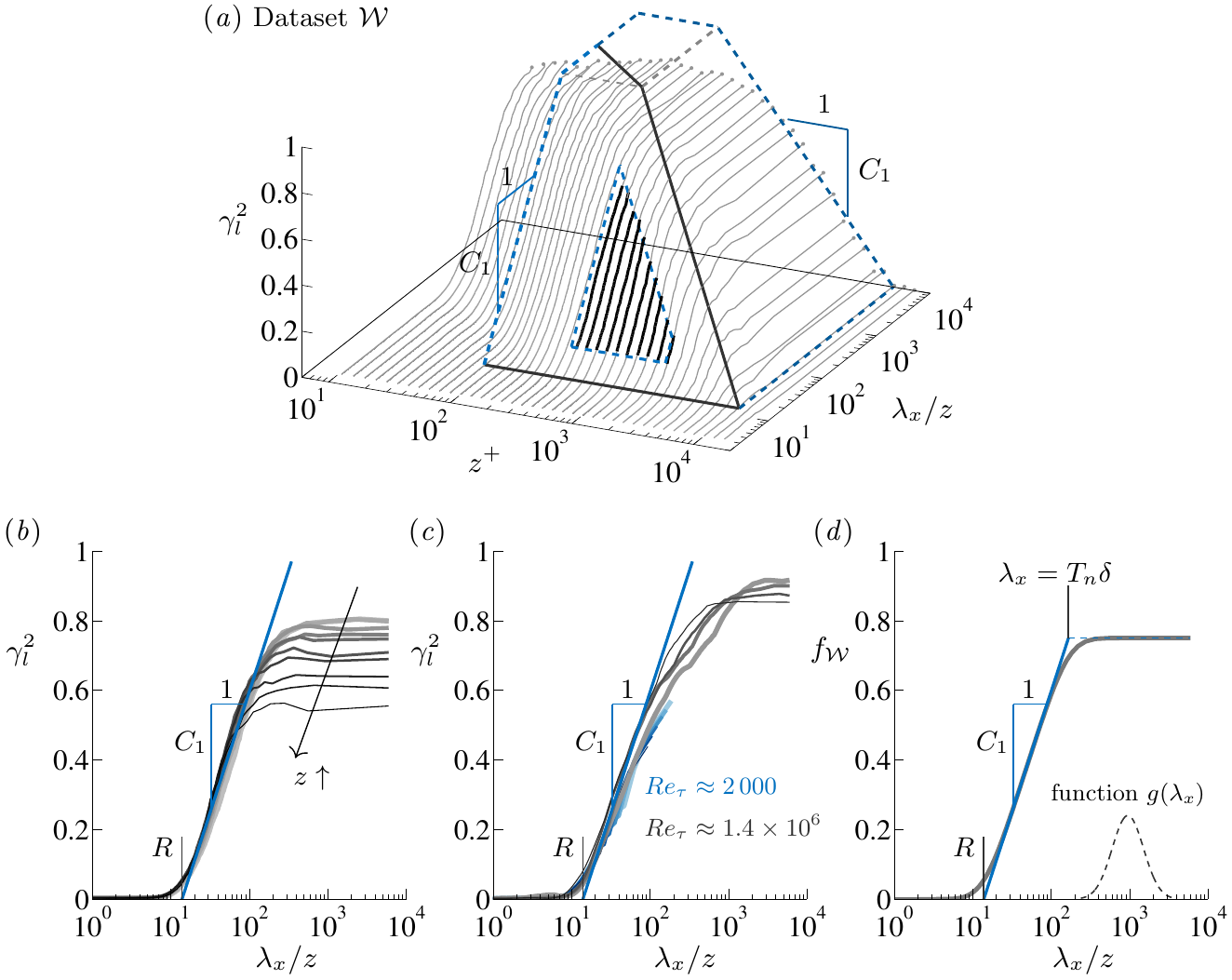}
   \caption{(\emph{a}) Coherence spectra of dataset $\mathcal{W}$ with the wavelength axis predivided by $z$. The thicker portions of the curves correspond to $3Re_\tau^{\nicefrac{1}{2}} < z^+ < 0.15Re_\tau$ and $20z < \lambda_x < 6\delta$. Note that the inclined plane with the solid boundary reflects (\ref{eq:Wfilt1}) with the constants stated in the text. (\emph{b-c}) Linear coherence spectra $\gamma^2_l(z,z_{\mathcal{W}};\lambda_x)$ for wall-normal positions within the range $3Re_\tau^{\nicefrac{1}{2}} < z^+ < 0.15Re_\tau$, for (\emph{b}) dataset $\mathcal{W}$ and (\emph{c}) the DNS and ASL data described in \S\,\ref{sec:Redata}. (\emph{d}) Filter following (\ref{eq:Wfilt2}) for $z^+ = 841$.} 
   \label{fig:Wfilt}
\end{figure}

\subsubsection{Universality of the data-driven filter}\label{sec:wallfil}
The $\gamma^2_l$ spectra of Figure~\ref{fig:LCSdataW} are re-plotted in Figure~\ref{fig:Wfilt}(\emph{a}), with a wall-scaling $\lambda_x/z$, and highlights the region where \citet{baars:2017a2} fitted (\ref{eq:baars}), resulting in constants $C_1 = 0.302$ and $R = 14.01$. Figure~\ref{fig:Wfilt}(\emph{b}) shows the spectra used for fitting, alongside (\ref{eq:baars}), while Figure~\ref{fig:Wfilt}(\emph{c}) displays the same trend line, superposed on the coherence spectra of the DNS and ASL data (\S\,\ref{sec:Redata}). Since (\ref{eq:baars}) describes the $\gamma^2_l$ growth for $Re_\tau \sim \mathcal{O}(10^3) - \mathcal{O}(10^6)$, it can be concluded that a wall-attached, self-similar structure is ingrained in $u$ (and that the constants in (\ref{eq:baars}) are universal). Note however that, for the ASL case, these constants depend on the temperature stratification \citep{krug:2019a}. From a fit to the $\mathcal{W}$ data, \citet{baars:2017a2} found that the large-scale limit of (\ref{eq:baars}) is $\lambda_x = T_n\delta$ (with $T_n \approx 10$), after which $\gamma^2_l$ transitions to a constant value for all larger wavelengths. Following (\ref{eq:dec}), (\ref{eq:baars}) and its region of validity, a data-driven spectral filter is formulated in (\ref{eq:Wfilt1}). This filter is in essence a model for the linear coherence spectrum and can be evaluated at any value of $Re_\tau$, due to its confirmed $Re_\tau$ universality.
\begin{eqnarray}
\label{eq:Wfilt1}
f^p_{\mathcal{W}}\left(z;\lambda_x\right) = 
     \begin{cases}
       0 & \quad \lambda_x < Rz \\
       {\rm min}\left\lbrace C_1\ln\left(\frac{\lambda_x}{z}\frac{1}{R}\right),\,1\right\rbrace & \quad Rz \leq \lambda_x \leq T_n\delta \\
       {\rm min}\left\lbrace C_1\ln\left(\frac{T_n\delta}{z}\frac{1}{R}\right),\,1\right\rbrace & \quad \lambda_x > T_n\delta \\
     \end{cases}
\end{eqnarray}
Subscript $\mathcal{W}$ denotes the wall-based reference on which this filter is based, whereas superscript $p$ refers to its piecewise nature. To obtain smooth transitions as in the data, a logarithmic convolution of (\ref{eq:Wfilt1}) with a log-normal distribution $g(\lambda_x)$ is performed:
\begin{eqnarray}
 \label{eq:Wfilt2}
 f_{\mathcal{W}}\left(z;\lambda_x\right) = f^p_{\mathcal{W}}\left(z;\lambda_x\right) *_l g\left(\lambda_x\right).
\end{eqnarray}
Here, $g(\lambda_x)$ spans six standard deviations, corresponding to 1.2 decades in $\lambda_x$, shown at the bottom right of Figure~\ref{fig:Wfilt}(\emph{d}) with a random amplitude. The displayed filter resembles the result for $z^+ = 841$. In summary, filter $f_{\mathcal{W}}\left(z;\lambda_x\right) \in [0,1]$ indicates the wavelength-dependent energy fraction, for a logarithmic region-position $z$, that is stochastically coherent with the near-wall region. Similarly, $(1-f_{\mathcal{W}})$ indicates the stochastically incoherent energy fraction. Coherent scales were characterised to appear at scales larger than $\lambda_x = Rz$ (constant $R$ refers to a characteristic aspect ratio), with a transition to a constant value at $\lambda_x = T_n\delta$ (constant $T_n$ is the nominal \emph{transition} scale). Only at sufficient scale separation, the filter saturates (\emph{i.e.} $f_W = 1$) at a scale smaller than $\lambda_x = T_n\delta$; this happens at $\lambda_x = T\delta$ (note that subscript $n$ is now dropped). Filter characteristics are summarized in Table~\ref{tab:constants}.

\subsubsection{Notes about the data-driven filter}\label{sec:Wnotes}
Filter (\ref{eq:Wfilt2}) is insensitive to the exact near-wall reference location. In fact, the ASL data employed friction velocity data ($z_R = 0$). Laboratory data at conditions similar to the $\mathcal{W}$ data, with a reference wall-shear stress sensor, yielded an indistinguishable coherence spectrogram \citep{baars:2017a2}. DNS data allows for a variation of $z_R$ and it has been confirmed that $\gamma^2_l$ spectra at $z^+ \apprge 80$ (relative to $z_R$) were virtually unaffected for $0 \leqslant z^+_R \apprle 15$. This insensitivity proves that our current filter-diagnostic for the wall-attached turbulence extracts inertial features in the logarithmic region that leave a distinct footprint throughout the near-wall region, as well as at the wall. 

Although the filter can saturate (an occurrence of perfect coherence), this state may only be reached asymptotically in actual coherence spectra. For instance, ASL coherence spectra in Figure~\ref{fig:Wfilt}(\emph{c}) do not reach unity at the largest scales ($\gamma^2_l \approx 0.92$ for the lowest ASL $z^+ \approx 3\,500$ spectrum, although filter $f_W$ saturates at $\lambda_x/z = R\exp(1/C_1) \approx 385$, corresponding to $\lambda_x/\delta \approx 385 z^+/Re_\tau \approx 0.96 < T_n\delta$). Many factors can cause the less-than-perfect coherence (as also pointed out in \S\,\ref{sec:wallcoh}) and is most likely related to an inherent 3D nature of the wall-attached structures and the continuous evolution of structures in the spatially developing TBL flow.

Figure~\ref{fig:LCSdataW} reveals an absence of coherence for $\lambda_x^+ \apprle 5\,000$. From now on it is assumed that (\ref{eq:Wfilt2}) is applicable to a logarithmic region starting at $z \sim \mathcal{O}(100\nu/U_\tau)$ \citep[see also][]{agostini:2017a}, because DNS data evidences a lower limit down to where the inner-spectral peak has a pronounced appearance in the spectrogram and integrated energy, say $z^+_T \approx 80$, see Figure~4 in \citet{baars:2017a2} and fully-resolved $\overline{u^2}$ profiles in \citet{samie:2018a}. Absence of any small-scale coherence in the $\mathcal{W}$ data is a topic for future work, but it is suspected that it is caused by experimental uncertainty in the spanwise alignment of the hot-wire probes at $z_R$ and $z$. 

\subsection{Coherence relative to a logarithmic-region reference}\label{sec:logregion}
\subsubsection{Coherence spectrogram and physical interpretation}\label{sec:logcoh}
\begin{figure} 
\vspace{10pt}
\centering
\includegraphics[width = 0.999\textwidth]{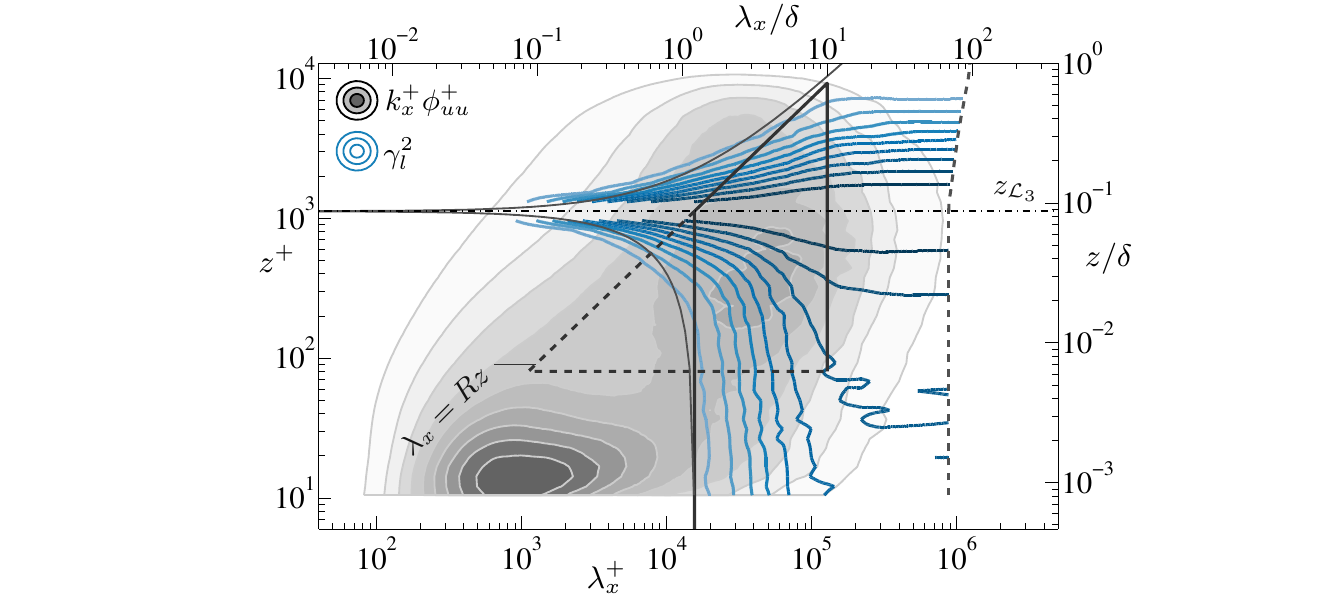}
   \caption{Dataset $\mathcal{L}_3$. Coherence spectrogram $\gamma^2_l(z,z_{\mathcal{L}_3};\lambda_x)$, relative to reference $z^+_{\mathcal{L}_3} \approx 1\,120$ (iso-contours 0.1:0.1:0.9), superposed on its associated premultiplied energy spectrogram $k^+_x\phi^+_{uu}$ (filled iso-contours 0.2:0.2:1.8), reproduced from Figure~\ref{fig:spintro1}.} 
   \label{fig:LCSdataL}
\end{figure}
Thus far the coherence trend has been discussed in the context of wall-attached motions. This section is concerned with $z_R$ residing in the logarithmic region (a location $z_{\mathcal{L}}$), from which a data-driven filter emerges that is dubbed the $\mathcal{L}$ filter. This filter separates the wall-attached motions into those that are coherent and incoherent with $z_{\mathcal{L}}$. 

\begin{figure} 
\vspace{10pt}
\centering
\includegraphics[width = 0.999\textwidth]{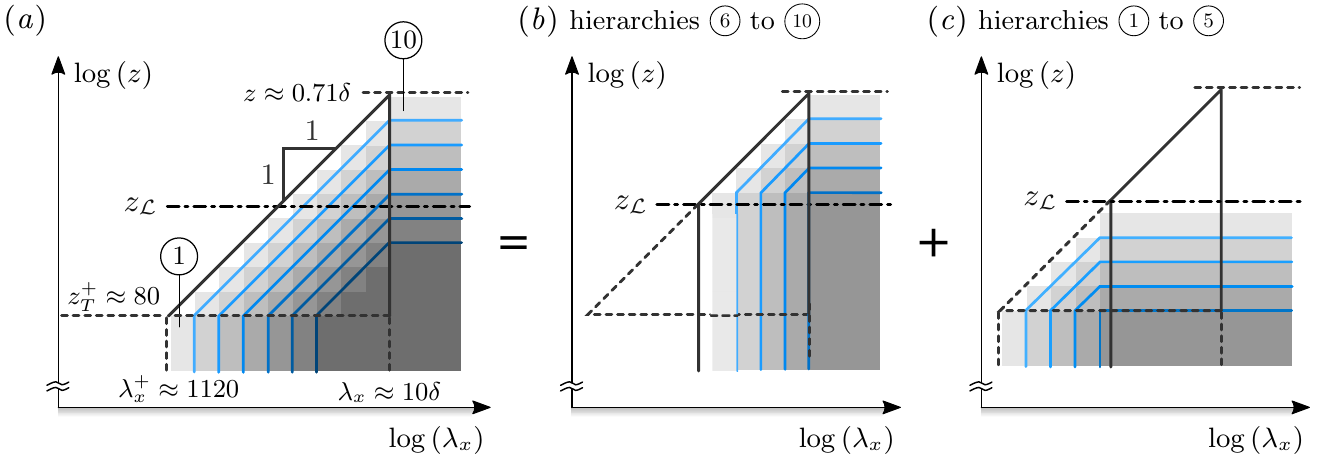}
   \caption{Sub-dividing the (\emph{a}) coherence spectrogram of wall-attached turbulence (Figure~\ref{fig:roadmapW}\emph{d}) into a component that is (\emph{b}) coherent and (\emph{c}) incoherent with a logarithmic-region reference $z_{\mathcal{L}}$.}
   \label{fig:roadmapL}
\end{figure}

A graph similar to Figure~\ref{fig:LCSdataW} is displayed in Figure~\ref{fig:LCSdataL} for the $\mathcal{L}_3$ data (see Figure~\ref{fig:LCSappA} in Appendix~\ref{sec:appA2} for all three datasets $\mathcal{L}_{1-3}$). To discuss the coherence in Figure~\ref{fig:LCSdataL} it is beneficial to recall the conceptually-reconstructed $\gamma^2_l$ contour in Figure~\ref{fig:roadmapW}(\emph{d}). When only the wall-attached turbulence is responsible for any coherence throughout the TBL, the coherence spectrogram can be sub-divided into two components (Figure~\ref{fig:roadmapL}). Ideally, for any given $z_{\mathcal{L}}$, only the wall-attached turbulence that extends beyond $z_{\mathcal{L}}$ is coherent with $z_{\mathcal{L}}$ (its associated sub-component is re-shown in Figure~\ref{fig:roadmapL}\emph{b}). Smaller wall-attached motions are not observed by $z_{\mathcal{L}}$ and do not contribute to the coherence relative to $z_{\mathcal{L}}$ (Figure~\ref{fig:roadmapL}\emph{c}). Figure~\ref{fig:LCSdataL} reflects the trend of Figure~\ref{fig:roadmapL}(\emph{b}). Note that the $\gamma^2_l$ contour is plotted with $\lambda_x \equiv U(z_{\mathcal{L}})/f$ for $z < z_{\mathcal{L}}$, since large-scale wall-attached structures (that are thus coherent with $z \geqslant z_{\mathcal{L}}$) comprise a convection velocity much larger than the mean velocity at locations in close proximity to the wall. For $z \ll z_{\mathcal{L}}$, say $z < z_{\mathcal{L}}/8$, the coherence spectra become non-zero for $\lambda_x \apprge Rz_{\mathcal{L}}$ (the vertical line extending down to the abscissa). For $z \gg z_{\mathcal{L}}$, coherence spectra tend towards the same trend as in Figure~\ref{fig:LCSdataW}.
When $z \sim z_{\mathcal{L}}$, coherence spectra reveal that even scales smaller than $\lambda_x < Rz$ become coherent with $z_{\mathcal{L}}$. This smaller-scale coherence is not associated with stochastically wall-attached motions, as there was no coherence in the $\lambda_x < Rz$ region with the near-wall reference $z_\mathcal{W}$ (Figure~\ref{fig:LCSdataW}). Turbulent scales for which $\gamma^2_l(z,z_{\mathcal{L}};\lambda_x) > \gamma^2_l(z,z_{\mathcal{W}};\lambda_x)$ are thus \emph{wall-detached} scales that are $z_{\mathcal{L}}$-coherent (Appendix~\ref{sec:appA2}). In \S\,\ref{sec:trdecom} it will become clear that for the purpose our work the small-scale coherence at $z \sim z_{\mathcal{L}}$ is irrelevant, as we work with the coherence model following Figures~\ref{fig:roadmapW} and~\ref{fig:roadmapL}.

\subsubsection{Universality of the data-driven filter}\label{sec:logfil}
\begin{figure} 
\centering
\includegraphics[width = 0.999\textwidth]{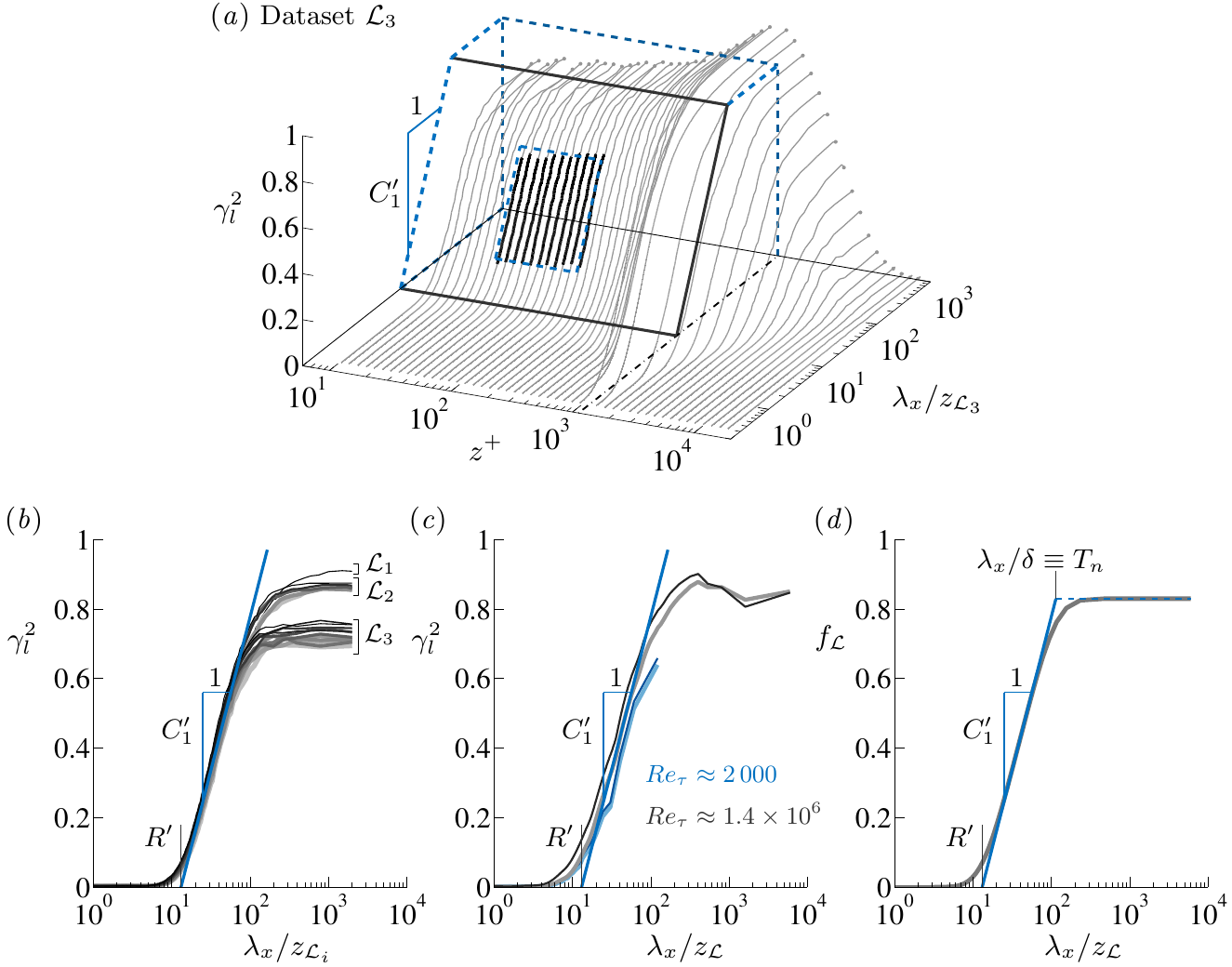}
   \caption{(\emph{a}) Coherence spectra of dataset $\mathcal{L}_3$ with the wavelength axis predivided by $z_{\mathcal{L}_3}$. The thicker portions of the curves correspond to $30 \apprle z^+ \apprle z_R^+/8$ and $20z_{\mathcal{L}_3} < \lambda_x < 5\delta$. Note that the inclined plane with the solid boundary reflects (\ref{eq:Lfilt1}) with the constants stated in the text. (\emph{b-c}) Linear coherence spectra $\gamma^2_l(z,z_{\mathcal{L}};\lambda_x)$ for wall-normal positions within the range $30 \apprle z^+ \apprle z^+_{\mathcal{L}}/8$, for (\emph{b}) datasets $\mathcal{L}_{1-3}$ and (\emph{c}) the DNS and ASL data. (\emph{d}) Filter following (\ref{eq:Lfilt2}) for $z^+_{\mathcal{L}} = 1\,120$.} 
   \label{fig:Lfilt}
\end{figure}
Since this work focuses on revealing spectral signatures associated with attached eddies, a filter relative to $z_{\mathcal{L}}$ is considered, with large enough $\Delta z$ so that only the wall-attached turbulence produces coherence. In other words, this spectral filter can decompose the \emph{wall-attached portion} of the turbulence at $z < z_{\mathcal{L}}$ into a $z_{\mathcal{L}}$-coherent and $z_{\mathcal{L}}$-incoherent component. Ideally, following Figure~\ref{fig:roadmapL}(\emph{b}), this filter equals $f^p_{\mathcal{W}}$ evaluated at $z = z_{\mathcal{L}}$ and is invariant with $z$ for $z < z_{\mathcal{L}}$. This can be confirmed using the data. Coherence spectra for $30 \apprle z^+ \apprle z_{\mathcal{L}}^+/8$ and datasets $\mathcal{L}_{1-3}$ show excellent collapse in Figure~\ref{fig:Lfilt}(\emph{b}). Precise criteria for the minimum $\Delta z$ are non-trivial and not the main focus of this paper. Following (\ref{eq:baars}) and \S\,\ref{sec:wallcoh}, fitting of 
\begin{eqnarray}
 \label{eq:baarsp}
 \gamma^2_l = C'_1\ln\left(\frac{\lambda_x}{z_\mathcal{L}}\frac{1}{R'}\right)
\end{eqnarray}
to the $\mathcal{L}_3$ data (highlighted portions of the profiles in Figure~\ref{fig:Lfilt}\emph{a}, described in the caption), yields $C'_1 = 0.3831$ and $R' = 13.18$. These constants are also seen to match the DNS and ASL data in Figure~\ref{fig:Lfilt}(\emph{b}), suggesting that (\ref{eq:baarsp}) is universal. Two DNS-data based coherence spectra with $z^+_{\mathcal{L}} \approx 192$ ($z_{\mathcal{L}}/\delta \approx 0.10$) and two ASL-data based spectra with $z^+_{\mathcal{L}} \approx 42\,300$ ($z_{\mathcal{L}}/\delta \approx 0.03$) are plotted, all of which satisfy $30 \apprle z^+ \apprle z^+_{\mathcal{L}}/6$. Note that, if the data adheres to the hypothesized underpinning in Figures~\ref{fig:roadmapW} and~\ref{fig:roadmapL}, the constants in (\ref{eq:baarsp}) and (\ref{eq:baars}) should be equal. The percentage variation of the constants (Table~\ref{tab:constants}) is attributed to the inherent simplification in using one convection velocity for all scales (wavelengths) and the---apparently different---contributions to the $z_{\mathcal{W}}$ and $z_{\mathcal{L}}$ coherence from \emph{non-self-similar} VLSMs (further discussed in \S\,\ref{sec:trdecom}). From \S\,\ref{sec:decom} it will become clear that the present variation of these constants does not have a major impact on the spectral energy decomposition, nor the conclusions of this work. A data-driven spectral filter (valid for $z < z_{\mathcal{L}}$) is now given as
\begin{eqnarray}
\label{eq:Lfilt1}
f^p_{\mathcal{L}}\left(z_{\mathcal{L}};\lambda_x\right) = 
     \begin{cases}
       0 & \quad \lambda_x < R'z_{\mathcal{L}} \\
       {\rm min}\left\lbrace C'_1\ln\left(\frac{\lambda_x}{z_{\mathcal{L}}}\frac{1}{R'}\right),\,1\right\rbrace & \quad R'z_{\mathcal{L}}\leq \lambda_x \leq T_n\delta \\
       {\rm min}\left\lbrace C'_1\ln\left(\frac{T_n\delta}{z_{\mathcal{L}}}\frac{1}{R'}\right),\,1\right\rbrace  & \quad \lambda_x > T_n\delta \\
     \end{cases}
\end{eqnarray}
Similar to (\ref{eq:Wfilt2}), a smooth filter is obtained with a logarithmic convolution:
\begin{eqnarray}
 \label{eq:Lfilt2}
 f_{\mathcal{L}}\left(z_{\mathcal{L}};\lambda_x\right) = f^p_{\mathcal{L}}\left(z_{\mathcal{L}};\lambda_x\right) *_l g\left(\lambda_x\right),
\end{eqnarray}
here $f_{\mathcal{L}} \in [0,1]$ and $f_{\mathcal{L}}$ is presented in Figure~\ref{fig:Lfilt}(\emph{d}). 

To conclude, two data-driven, universal spectral filters have been presented: filter $f_{\mathcal{W}}$, for extracting the spectral energy in the TBL that is stochastically coherent with the wall (\emph{e.g.} wall-attached), and $f_{\mathcal{L}}$, which reveals the sub-fraction of the wall-attached energy that is stochastically coherent with a position in the logarithmic region, $z_{\mathcal{L}}$.
\begin{table}
  \begin{center}
  \vspace*{-6pt}
  \begin{tabular}{ccccccc}
  filter & \multicolumn{3}{c}{Constants} & Fitted to & Equations & Region\\ \hline
  $f_\mathcal{W}(z;\lambda_x)$	& $C_1 = 0.3017$ & $R = 14.01$ & $T_n = 10$ & dataset $\mathcal{W}$ & (\ref{eq:Wfilt1}) -- (\ref{eq:Wfilt2}) & $z^+ > z^+_T$ \\[1.2pt]
  $f_\mathcal{L}(z;\lambda_x)$	& $C'_1 = 0.3831$ & $R' = 13.18$ & $T_n = 10$ & dataset $\mathcal{L}_3$ & (\ref{eq:Lfilt1}) -- (\ref{eq:Lfilt2}) & $z < z_{\mathcal{L}}$\\ \hline
  diff. $\mathcal{L}$ vs. $\mathcal{W}$ & +27\,\% & -6\,\% & ~ & ~ & ~ & ~\\
  \end{tabular}
  \caption{Constants of the two data-driven spectral filters. Nominally $z^+_T = 80$.}
  \label{tab:constants}
  \end{center}
\end{table}

\section{Decomposing the streamwise energy spectra}\label{sec:decom}

\subsection{Triple-decomposition via data-driven filters}\label{sec:trdecom}
\begin{figure} 
\vspace{10pt}
\centering
\includegraphics[width = 0.999\textwidth]{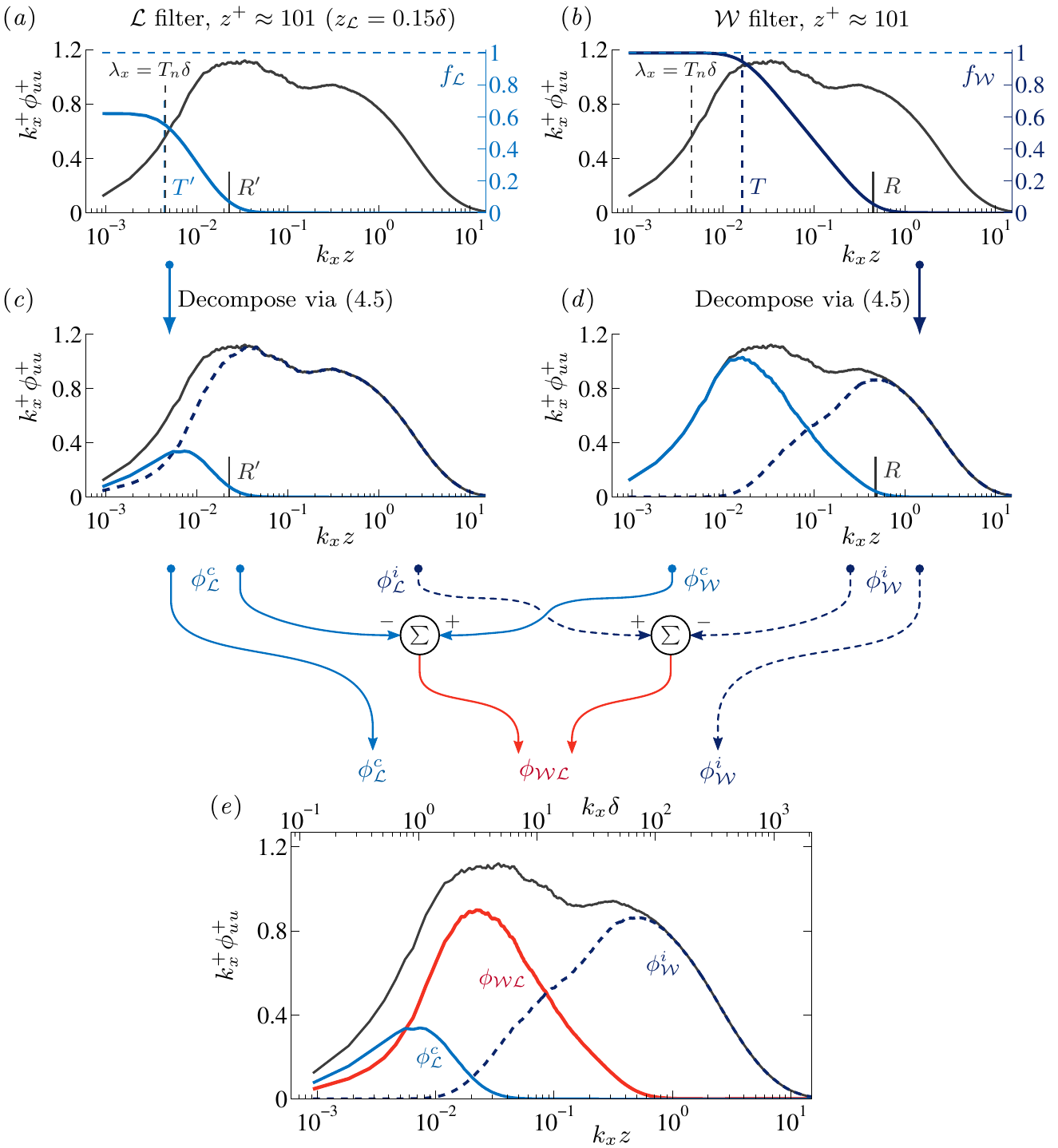}
   \caption{Diagram of the triple decomposition of the energy spectrum at $z^+ \approx 101$ ($\mathcal{W}$ data). (a,b) $\mathcal{L}$ and $\mathcal{W}$ filters overlaid on the energy spectrum. (c,d) Coherent (solid line) and incoherent (dashed line) components of the energy spectrum, via the $\mathcal{L}$ and $\mathcal{W}$ filters, respectively. (e) Resulting three spectral sub-components.} 
   \label{fig:trdecom}
\end{figure}

How the streamwise energy spectra are decomposed with the use of data-driven filters $f_{\mathcal{W}}$ and $f_{\mathcal{L}}$ is outlined in Figure~\ref{fig:trdecom}. A \emph{triple decomposition} is performed, so that a single energy spectrum is split into three sub-components, each of which are interpreted in the context of the $\mathcal{A}$, $\mathcal{B}$ and $\mathcal{C}$ components of Perry \etal~(Figure~\ref{fig:spperry}). 

In Figure~\ref{fig:trdecom}(\emph{b}), $f_{\mathcal{W}}$ is overlaid on the energy spectrum. At the high-wavenumber end the filter is bounded by $k_xz = 2\pi/R \approx 0.45$. At low wavenumbers, the filter-amplitude reaches 1 at scale $T$. Figure~\ref{fig:trdecom}(\emph{a}) is similar to Figure~\ref{fig:trdecom}(\emph{b}) but presents $f_{\mathcal{L}}$ with $z_{\mathcal{L}} = 0.15\delta$. The filter is bounded by $k_xz = 2\pi/R'(z/z_L) \approx 0.023$) and transitions at scale $T'$ corresponding to $\lambda_x = T_n\delta$ (and plateaus to a value less than one). Figures~\ref{fig:trdecom}(\emph{c}) and~\ref{fig:trdecom}(\emph{d}) show the coherent (solid line) and incoherent (dashed) components of the energy spectrum, per (\ref{eq:dec}). Formulations (\ref{eq:trdecom1})--(\ref{eq:trdecom3}) define the three spectral sub-components, which yield the total spectrum when added: $\phi_{uu} = \phi_{\mathcal{L}}^c + \phi_{\mathcal{W}\mathcal{L}} + \phi_{\mathcal{W}}^i$.
\begin{eqnarray}
 \label{eq:trdecom1}
 \phi_{\mathcal{L}}^c \equiv &\phi_{uu}&f_{\mathcal{L}} \\
 \label{eq:trdecom2}
 \phi_{\mathcal{W}}^i \equiv &\phi_{uu}&\left(1-f_{\mathcal{W}}\right) \\
 \label{eq:trdecom3}
 \phi_{\mathcal{W}\mathcal{L}} \equiv &\phi_{uu}&\left(f_{\mathcal{W}} - f_{\mathcal{L}}\right)
\end{eqnarray}
A diagram leading to Figure~\ref{fig:trdecom}(\emph{e}) demonstrates these definitions:\\[-8pt]
\begin{enumerate}[labelwidth=0.65cm,labelindent=0pt,leftmargin=0.65cm,label=(\roman*),align=left]
\item \noindent At high wavenumbers, $\phi_{\mathcal{W}}^i$ is taken as the $f_{\mathcal{W}}$-based incoherent component. This stochastically incoherent energy cannot be estimated using a LSE procedure with a wall-based input. Following the classification of Perry \etal, the underlying turbulence of such a component may be thought of as the small-scale type $\mathcal{C}$ eddies. However, it can also include any (stochastically) detached (non)-self-similar motions, such as wall-incoherent VLSMs.\\[-10pt]
\item \noindent At low wavenumbers, $\phi_{\mathcal{L}}^c$ is taken as the energy that is coherent via $f_{\mathcal{L}}$. This large-scale wall-attached energy is also coherent with $z_{\mathcal{L}} = 0.15\delta$. Physically, this wall-attached component may include self-similar structures reaching beyond $z_{\mathcal{L}}$ and non-self-similar structures that are coherent with $z_{\mathcal{L}}$ (\emph{e.g.} VLSMs).\\[-10pt]
\item \noindent A remaining component is dubbed $\phi_{\mathcal{W}\mathcal{L}}$. Its energy equals the wall-coherent energy via $f_{\mathcal{W}}$, $\phi_{\mathcal{W}}^c$, minus its fraction that is coherent with $z_{\mathcal{L}}$, being $\phi_{\mathcal{L}}^c$ (or equivalently, all $z_{\mathcal{L}}$-incoherent energy, $\phi_{\mathcal{L}}^i$, minus the wall-incoherent energy, $\phi_{\mathcal{W}}^i$). On account of the above, $\phi_{\mathcal{W}\mathcal{L}}$ is the wall-coherent energy that resides below $z_{\mathcal{L}}$ and can include both self-similar and non-self-similar components.\\[-8pt]
\end{enumerate}

\begin{figure} 
\vspace{10pt}
\centering
\includegraphics[width = 0.999\textwidth]{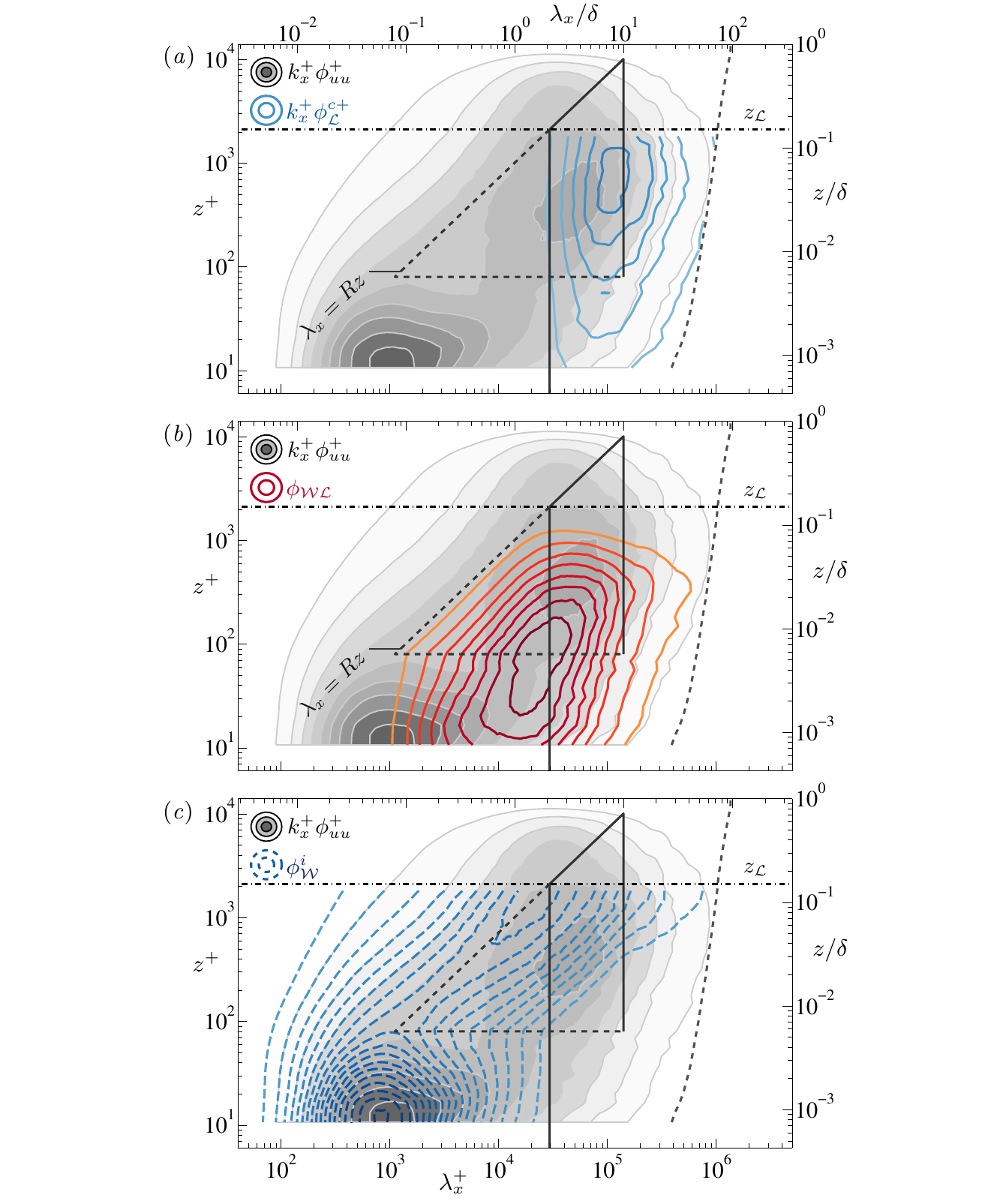}
   \caption{Dataset $\mathcal{W}$ with $Re_\tau \approx 14\,100$. (a-c) Premultiplied energy spectrograms of the three spectral sub-components (for $z < z_{\mathcal{L}}$), each of them overlaid on the total energy spectrogram (filled iso-contours 0.2:0.2:1.8). Iso-contours corresponding to the three sub-components are shown with a twice-as-fine level spacing, 0.1:0.1:1.8.} 
   \label{fig:trdW1}
\end{figure}

\begin{figure} 
\vspace{10pt}
\centering
\includegraphics[width = 0.999\textwidth]{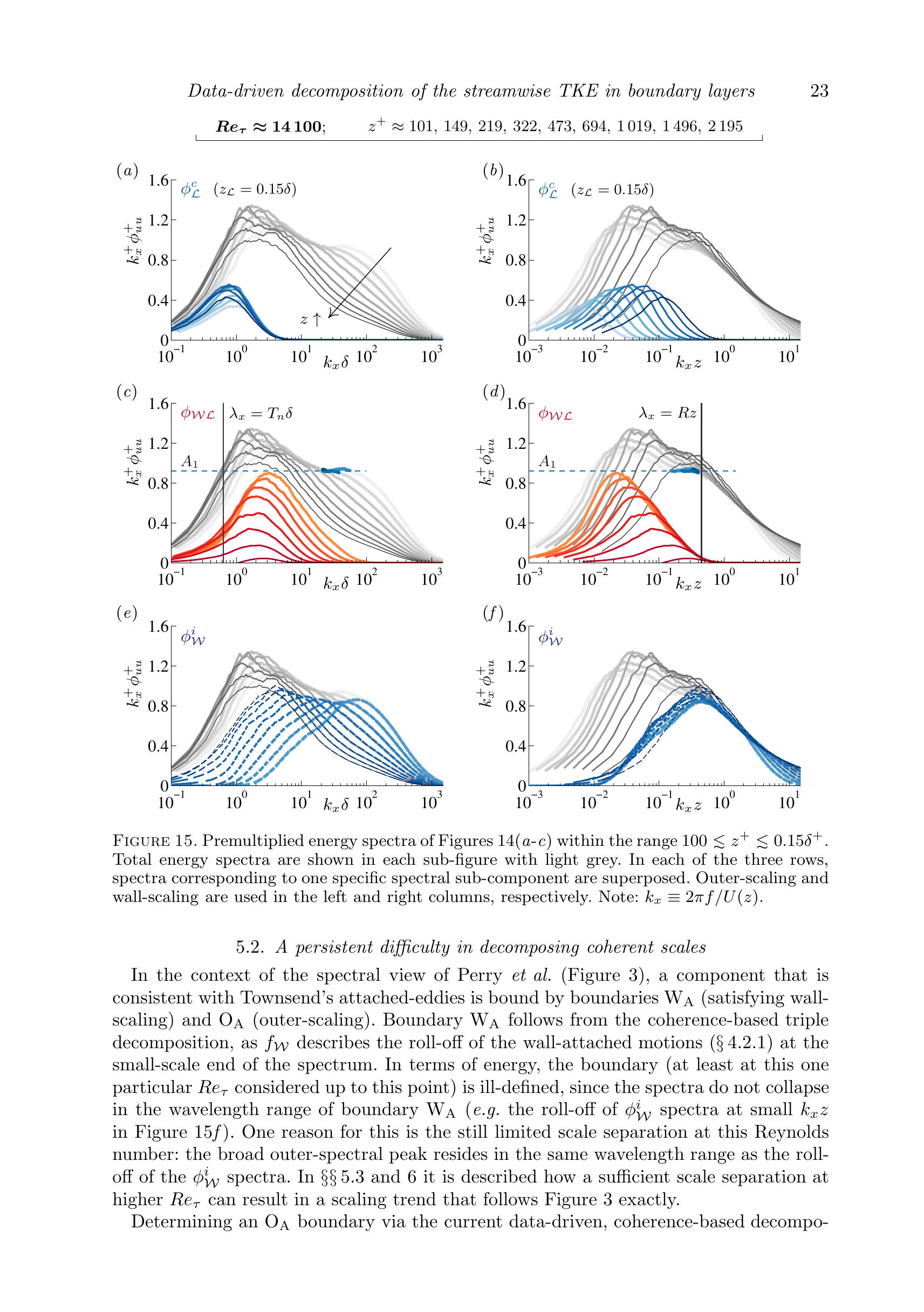}
   \caption{Premultiplied energy spectra of Figures~\ref{fig:trdW1}(\emph{a}-\emph{c}) within the range $100 \apprle z^+ \apprle 0.15\delta^+$. Total energy spectra are shown in each sub-figure with light grey. In each of the three rows, spectra corresponding to one specific spectral sub-component are superposed. Outer-scaling and wall-scaling are used in the left and right columns, respectively. Note: $k_x \equiv 2\pi f/U(z)$.}
   \label{fig:trdW2}
\end{figure}

After performing the aforementioned decomposition for all $z$ we can overlay the three resulting energy spectrograms on the total energy spectrogram in Figures~\ref{fig:trdW1}(\emph{a}-\emph{c}). In addition, all spectral components within the range $100 \apprle z^+ \apprle 0.15\delta^+$ are plotted with wall-scaling and outer-scaling in Figure~\ref{fig:trdW2}. For Figure~\ref{fig:trdW1}, sub-components from the triple decomposition are only considered for $z < z_{\mathcal{L}}$. In the near-wall region, here $z^+ \apprle z^+_T$ (with nominally $z^+_T = 80$), $f_{\mathcal{W}}$ is $z$-invariant and taken as $f_{\mathcal{W}}(z^+_T;\lambda_x)$. As this work is concerned with the spectral structure in the logarithmic region, the spectral sub-components within the near-wall region are not considered further.

Component $\phi_{\mathcal{W}}^i$ follows the unfiltered spectra at the smallest energetic scales. It is only at a wall-scale of about $\lambda_x = Rz$ that $f_{\mathcal{W}}$ becomes active. The $\phi_{\mathcal{W}}^i$ energy roll-off at the large-scale end (small $k_x$ in Figure~\ref{fig:trdW2}\emph{f}) is generated by the product of the total spectrum and the ramp of $(1-f_{\mathcal{W}})$ within $Rz \apprle \lambda_x \apprle T_n\delta$. Since the filter ramp is universal in wall-scaling, and the measured spectra do not plateau in that region, the $\phi_{\mathcal{W}}^i$ spectra do not obey an unambiguous wall-scaling (Figure~\ref{fig:trdW2}\emph{f}). Maxima reside at $\lambda_x \approx Rz$, simply because the roll-off induced by the filter is steep enough to suppress any spectral increase at the large-scale end. Component $\phi_{\mathcal{W}\mathcal{L}}$ bounds $\phi_{\mathcal{W}}^i$ and its ramp-up at the small-scale end shows a reasonable collapse in wall-scaling (Figure~\ref{fig:trdW2}\emph{d}). Amplitudes of its peak decay rapidly with increasing $z$ position, due to $f_{\mathcal{L}}$ taking effect at an outer-scaling of $\lambda_x \approx R'z_{\mathcal{L}}$ (and due to the limited Reynolds number). An outer-scaling roll-off at the large-scale end of the spectrum is governed by the universal $f_{\mathcal{L}}$ ramp, multiplied by $\phi_{uu}$. Because the measured spectra, $\phi_{uu}$, do not obey perfect outer-scaling at the large wavelengths (with $z$ variation), the $\phi_{\mathcal{W}\mathcal{L}}$ spectra do not either. Finally, component $\phi_{\mathcal{L}}^c$ does comprise a certain degree of outer-scaling collapse at its small-scale end where $f_{\mathcal{L}}$ dictates its decay (Figure~\ref{fig:trdW2}\emph{a}). Before proceeding with an in-depth discussion on the spectral components, it is instructive to first concentrate on the limitations of the triple decomposition and the restricted scale separation at finite $Re_\tau$.

\subsection{A persistent difficulty in decomposing coherent scales}\label{sec:decomcoh}
\begin{figure} 
\vspace{10pt}
\centering
\includegraphics[width = 0.999\textwidth]{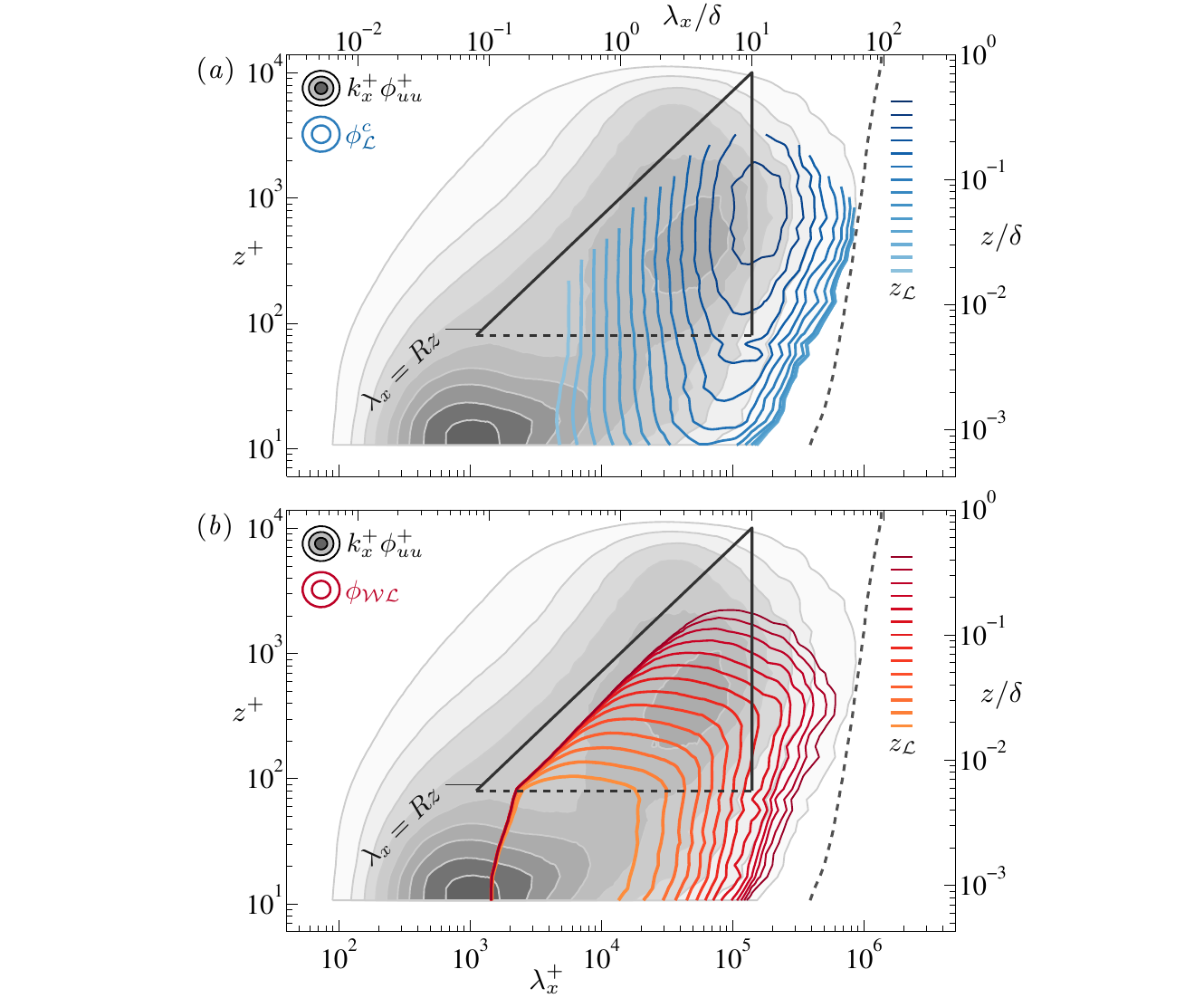}
   \caption{Dataset $\mathcal{W}$ with $Re_\tau \approx 14\,100$. (\emph{a},\emph{b}) Similar to Figures~\ref{fig:trdW1}(\emph{a},\emph{b}) but only with the 0.2 iso-contour shown for the premultiplied energy spectrograms of the spectral sub-components $\phi_{\mathcal{L}}^c$ and $\phi_{\mathcal{W}\mathcal{L}}$ (for $z < z_{\mathcal{L}}$). Different iso-contours correspond to a range of $z_{\mathcal{L}}$; their locations $z = z_{\mathcal{L}}$ are shown with the horizontal lines on the right-hand side. Note: $\lambda_x \equiv U(z)/f$.}
   \label{fig:trdW3}
\end{figure}

In the context of the spectral view of Perry \etal~(Figure~\ref{fig:spperry}), a component that is consistent with Townsend's attached-eddies is bound by boundaries W$_{\rm A}$ (satisfying wall-scaling) and O$_{\rm A}$ (outer-scaling). Boundary W$_{\rm A}$ follows from the coherence-based triple decomposition, as $f_{\mathcal{W}}$ describes the roll-off of the wall-attached motions (\S\,\ref{sec:wallcoh}) at the small-scale end of the spectrum. In terms of energy, the boundary (at least at this one particular $Re_\tau$ considered up to this point) is ill-defined, since the spectra do not collapse in the wavelength range of boundary W$_{\rm A}$ (\emph{e.g.} the roll-off of $\phi^i_\mathcal{W}$ spectra at small $k_xz$ in Figure~\ref{fig:trdW2}\emph{f}). One reason for this is the still limited scale separation at this Reynolds number: the broad outer-spectral peak resides in the same wavelength range as the roll-off of the $\phi^i_\mathcal{W}$ spectra. In \S\S\,\ref{sec:kmin1} and~\ref{sec:Retrend} it is described how a sufficient scale separation at higher $Re_\tau$ can result in a scaling trend that follows Figure~\ref{fig:spperry} exactly.

Determining an O$_{\rm A}$ boundary via the current data-driven, coherence-based decomposition technique remains difficult. Location $z_{\mathcal{L}}$ is subject to choice, and irrespective of $z_{\mathcal{L}}$, the wall-attached (coherent) energy does supposedly include both self-similar and non-self-similar scales at all of its energetic wavelengths, meaning that a distinction between (1) self-similar wall-attached scales (envisioned as the $\mathcal{A}$ energy), and (2) non-self-similar wall-attached scales (categorized in $\mathcal{B}$) is impossible via the current technique. While the conceptual reconstruction of the wall-coherence in Figure~\ref{fig:roadmapW}---with the adhering contours from the data---does imply a self-similar range of scales ($\AR = R \approx 14$), the absolute energy in the wall-coherent region ($\lambda_x > Rz$) does include non-self-similar contributions, simply because the outer-spectral peak in the boundary layer spectrogram resides in that region and does not strictly adhere to simultaneous wall-scaling and outer-scaling. Choosing a small $z_{\mathcal{L}}$ does likely result in $\phi_{\mathcal{W}\mathcal{L}}$ containing lesser non-self-similar energy. But, an inherent drawback of a small $z_{\mathcal{L}}$, is the limited range of self-similar wall-attached scales in $\phi_{\mathcal{W}\mathcal{L}}$ (namely scales reaching up to $z < z_{\mathcal{L}}$). Figures~\ref{fig:trdW3}(\emph{a},\emph{b}) visualizes how a different $z_{\mathcal{L}}$ re-distributes the energy between the $\phi_{\mathcal{L}}^c$ and $\phi_{\mathcal{W}\mathcal{L}}$ contributions (component $\phi_{\mathcal{W}}^i$ is unaffected). When moving up in $z_{\mathcal{L}}$, more wall-attached energy is assigned to $\phi_{\mathcal{W}\mathcal{L}}$, per the expense of what ends up in $\phi_{\mathcal{L}}^c$. Since $f_{\mathcal{L}}$ only fully matures when $z_{\mathcal{L}} < T_n\delta/\exp(1/C'_1)/R' \approx 0.056\delta$ ($f_{\mathcal{L}}$ plateaus at 1 in Figure~\ref{fig:Lfilt}\emph{d}), $\phi_{\mathcal{W}\mathcal{L}}$ will certainly contain the non-self-similar contributions from VLSM-type structures for $z > 0.056\delta$. 

\subsection{When can we possibly observe a $k_x^{-1}$ scaling region?}\label{sec:kmin1}
\begin{figure} 
\vspace{10pt}
\centering
\includegraphics[width = 0.999\textwidth]{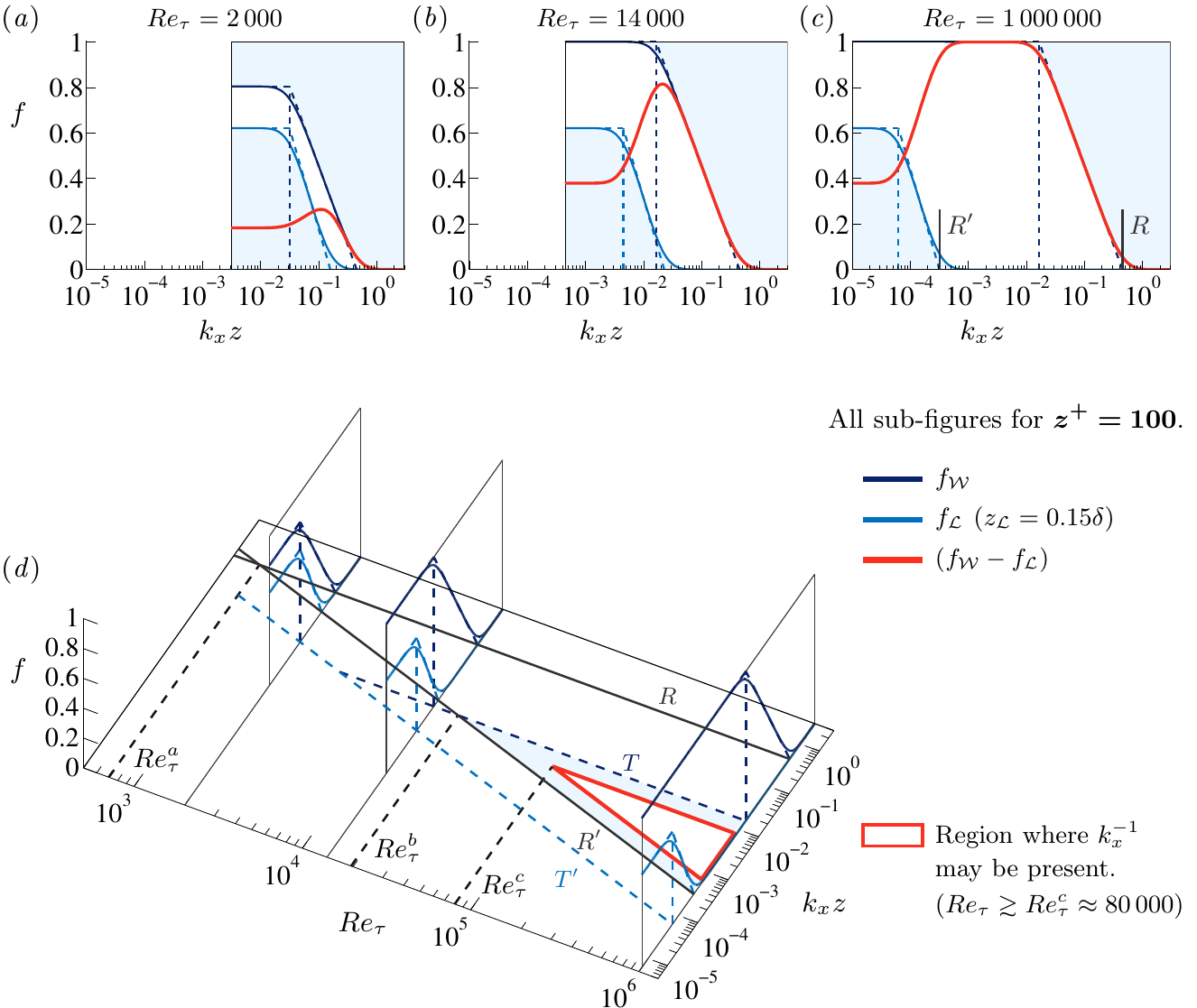}
   \caption{(\emph{a}-\emph{c}) Filters $f_{\mathcal{W}}$ and $f_{\mathcal{L}}$ for $z^+ = 100$ and $z_{\mathcal{L}} = 0.15\delta$ at three different $Re_\tau$. The low-wavenumber boundaries of the curves on each plot correspond to $\lambda_x = 100\delta$. (\emph{d}) Variation of the filters with $Re_\tau$, including the filters' onsets ($R$ and $R'$) and transitions ($T$ and $T'$).} 
   \label{fig:k1pres}
\end{figure}

We consider the Reynolds number dependence of the data-driven filters, in order to appraise the required scale separation for possibly observing a $k_x^{-1}$ region. Filters $f_{\mathcal{W}}$ and $f_{\mathcal{L}}$ are shown in Figures~\ref{fig:k1pres}(\emph{a}-\emph{b}) for $z^+ = 100$ at three different $Re_\tau$ values, closely resembling the DNS and ASL data (\S\,\ref{sec:Redata}), as well as the two-point laboratory data (\S\,\ref{sec:twopoint}). All filter curves start at $\lambda_x = 100\delta$. Following the triple decomposition of \S\,\ref{sec:trdecom}, the shaded area at the small wavenumber-end, cornered by $f_{\mathcal{L}}$ (here $z_{\mathcal{L}} = 0.15\delta$), represents the fraction of $k_x\phi_{uu}$ that forms $\phi_{\mathcal{L}}^c$, while the shaded area bounded by $f_{\mathcal{W}}$ at the high wavenumber-end reflects $\phi_{\mathcal{W}}^i$. The unshaded area in between the filter curves equals the fraction of spectral energy that forms $\phi_{\mathcal{W}\mathcal{L}}$, drawn with the black curves. Figure~\ref{fig:k1pres}(\emph{d}) presents Figures~\ref{fig:k1pres}(\emph{a}-\emph{c}) in the $Re_\tau$-continuum as three planes in $(f,Re_\tau,k_xz)$--space, and has indicated the footprint of the filters in the $(Re_\tau,k_xz)$--plane.

Both filters transition to a constant value (a plateau) at wavelengths larger than $\lambda_x = T_n\delta$, unless their amplitude reaches 1 at smaller scales. With increasing $Re_\tau$, the plateau-amplitude of $f_{\mathcal{L}}$ grows as $f_{\mathcal{L}}$ starts at a fixed inner-scaling (boundary $R$). At $Re_\tau = (z^+/T_n)R\exp(1/C_1) \approx 3\,850$ the filter plateau reaches unity, corresponding to the point in Figure~\ref{fig:k1pres}(\emph{d}) where the $T'$ and $T$ boundaries separate. Boundary $T$ now becomes fixed in inner-scaling, while $T'$ remains fixed in outer-scaling, because the onset of $f_{\mathcal{L}}$ is also outer-scaled, \emph{e.g.} $\lambda_x = R'z_{\mathcal{L}} = 0.15R'\delta$. At the low $Re_\tau$ regime, $R$ and $R'$ intersect at $Re_\tau^a \equiv z^+/(z_{\mathcal{L}}/\delta)\cdot(R/R') \approx 708$, meaning that $\phi_{\mathcal{W}\mathcal{L}}$ becomes non-existent (\emph{e.g.} there is no wall-coherent energy at $z^+ = 100$ that is incoherent with $z_{\mathcal{L}} = 0.15\delta$, simply because these positions merge: $z^+_{\mathcal{L}} = 0.15Re_\tau^a \approx 100$. At higher $Re_\tau$, specifically at $Re_\tau^b \equiv z^+/(z_{\mathcal{L}}/\delta)\exp(1/C_1)R/R' \approx 19\,500$, $f_{\mathcal{L}}$ only becomes active in a scale range where $f_{\mathcal{W}}$ has fully saturated. Though, because of the smooth filter transitions this does not happen until $Re_\tau^c \approx 80\,000$. And so, for $Re_\tau > Re_\tau^c$ (at $z^+ = 100$ and with $z_{\mathcal{L}} = 0.15\delta$), a range of scales starts to appear where all energy of $k_x\phi_{uu}$ would be assigned to $\phi_{\mathcal{W}\mathcal{L}}$ (only at $Re_\tau \approx 10^6$ this region spans one decade of scales). Moving from coherence to energy spectra, ideally, the scales defining this region do not comprise any wall-incoherent energy, nor do they comprise energy of structures that are coherent with $z > z_{\mathcal{L}}$. Thus, if the energy in this region is pre-dominantly induced by \emph{self-similar} attached eddies, this is the region where a $k_x^{-1}$ could be present. The current $Re_\tau$ estimate at which $\phi_{uu} \propto k_x^{-1}$ may be observed agrees reasonably well with the study by \citet{chandran:2017a}, where they predicted that an appreciable $k^{-1}$ scaling region can only appear for $Re_\tau \apprge 60\,000$ (from examination of 2D, streamwise-spanwise $u$ spectra). Note that the aforementioned analysis holds for $z^+ = 100$ and that a $k_x^{-1}$ region is expected to shrink linearly with $z$, meaning that $Re_\tau \sim \mathcal{O}(10^6)$ is required for a $\phi_{uu} \propto k_x^{-1}$ at $z^+ = 1000$.

\section{Reynolds number variation of the decomposed energy spectra}\label{sec:Retrend}

\subsection{Spectrograms, spectra and scaling behaviours}\label{sec:Respectra}
\begin{figure} 
\vspace{10pt}
\centering
\includegraphics[width = 0.999\textwidth]{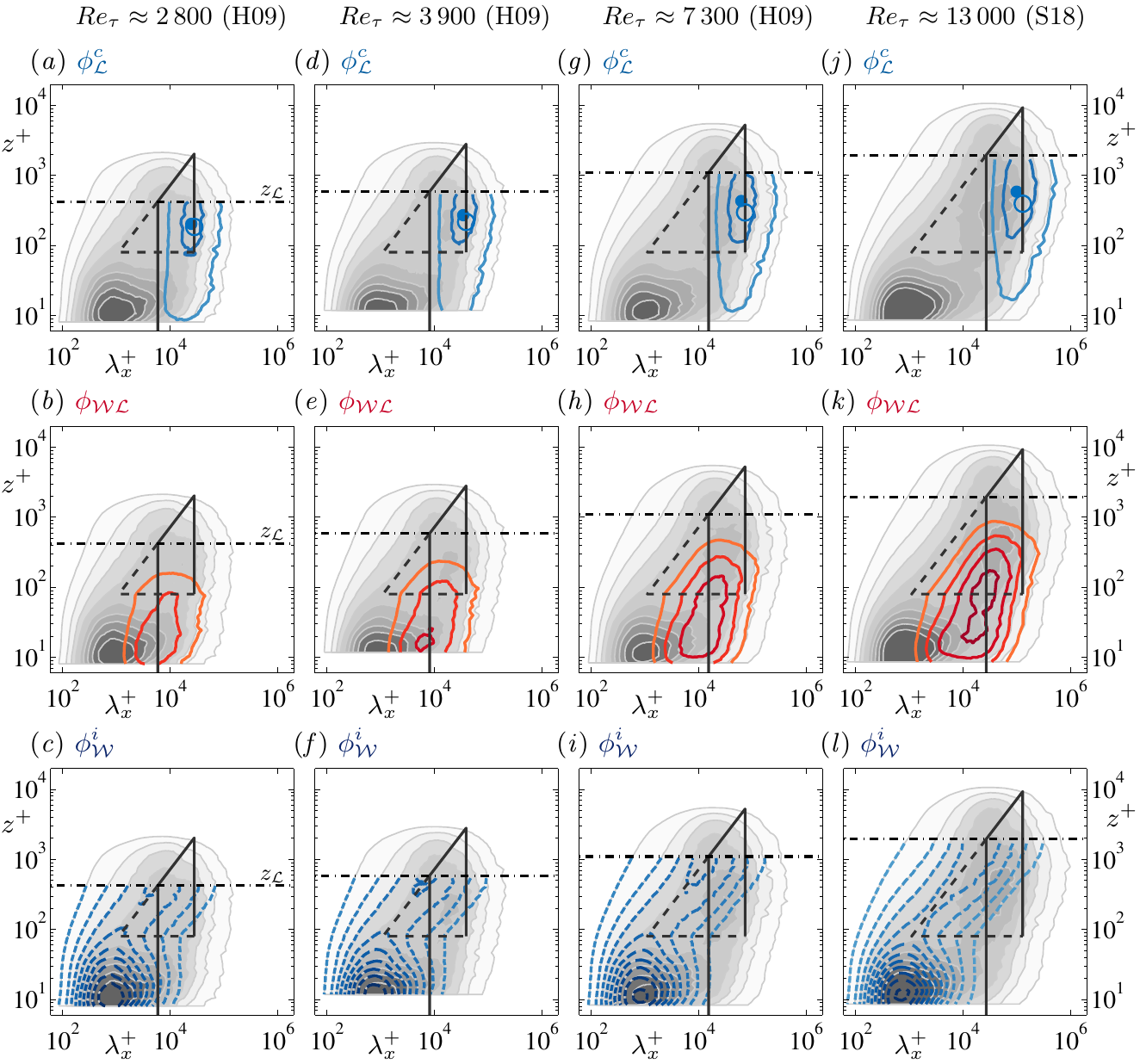}
   \caption{Each column (per $Re_\tau$ quoted on top) is similar to Figure~\ref{fig:trdW1}: premultiplied energy spectrograms of the three spectral sub-components (for $z < z_{\mathcal{L}}$, iso-contours 0.2:0.2:1.8), each of them superposed on the total energy spectrogram (filled iso-contours 0.2:0.2:1.8). Filter $f_{\mathcal{L}}$, with $z_{\mathcal{L}} = 0.15\delta$, is applied for $z \apprle z_{\mathcal{L}}$; $f_{\mathcal{W}}$ is applied for $z^+_T \apprle z^+ \apprle z^+_{\mathcal{L}}$, while $f_{\mathcal{W}}$ is $z$-invariant for $z^+ \apprle z^+_T$ and equal to $f_{\mathcal{W}}(z^+_T;\lambda_x)$. Data are from H09: \citet{hutchins:2009a} and S18: \citet{samie:2018a}, see \S\,\ref{sec:Redata}. Note: $\lambda_x \equiv U(z)/f$.}
   \label{fig:Retrend1}
\end{figure}
\begin{figure} 
\vspace{10pt}
\centering
\includegraphics[width = 0.999\textwidth]{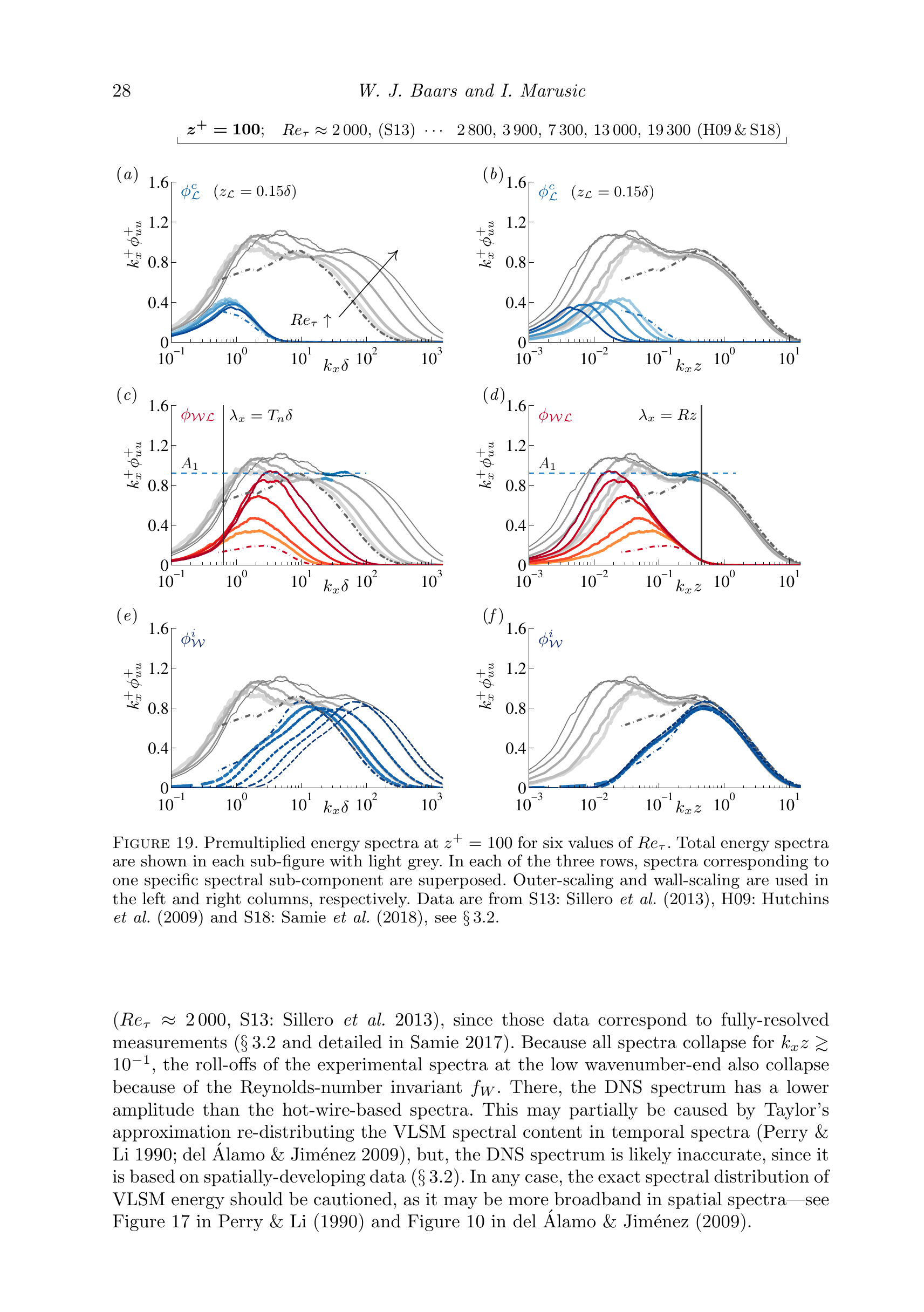}
   \caption{Premultiplied energy spectra at $z^+ = 100$ for six values of $Re_\tau$. Total energy spectra are shown in each sub-figure with light grey. In each of the three rows, spectra corresponding to one specific spectral sub-component are superposed. Outer-scaling and wall-scaling are used in the left and right columns, respectively. Data are from S13: \citet{sillero:2013a}, H09: \citet{hutchins:2009a} and S18: \citet{samie:2018a}, see \S\,\ref{sec:Redata}.} 
   \label{fig:Retrend2}
\end{figure}

Reynolds number trends of the triple-decomposed spectrograms are presented in Figure~\ref{fig:Retrend1} from singe-point data at four Reynolds numbers, spanning $Re_\tau \approx 2\,800$ to $13\,000$. Similarly as in Figure~\ref{fig:trdW1}, $z_{\mathcal{L}} = 0.15\delta$ and the triangles indicate wall-scaling $\lambda_x = Rz$ and outer-scaling $\lambda_x = T_n\delta$. Spectra at $z^+ = 100$ (interpolated from the spectrograms) are detailed in Figure~\ref{fig:Retrend2} for all five single-point hot-wire datasets (\S\,\ref{sec:Redata}). Additionally, a spatial spectrum from the DNS data is superposed. Components $\phi_{\mathcal{L}}^c$ and $\phi_{\mathcal{W}}^i$ are now discussed, after which we focus on $\phi_{\mathcal{W}\mathcal{L}}$ in \S\,\ref{sec:Remature}.

Large-scale contribution $\phi_{\mathcal{L}}^c$ is roughly Reynolds-number independent at $z^+ = 100$ (Figure~\ref{fig:Retrend2}\emph{a}). A minor decrease of energy within the largest wavelengths, with increasing $Re_\tau$, is an artifact of using a single convection velocity to generate $k_x \equiv 2\pi f/U_c$, here $U_c = U(z^+ = 100)$. Since the $\phi_{\mathcal{L}}^c$ component is associated with strong coherence in $z$, from the wall up to at least $z_{\mathcal{L}}$, these structures make up global modes \citep{bullock:1978a,delalamo:2003a} that are associated with an outer-scaled convection velocity \citep{delalamo:2004a}. Hence, when the local mean velocity at $z_{\mathcal{L}} = 0.15\delta$ is employed for each Reynolds-number dataset to construct the scale-axis in Figure~\ref{fig:Retrend2}(\emph{a}), the large-scale end of the spectra collapse \citep[see Appendix~\ref{sec:appA3} for further details, as well as][]{samie:2017phd}. A wall-normal trend of $\phi_{\mathcal{L}}^c$, together with its $Re_\tau$ dependence, is apparent from Figure~\ref{fig:Retrend1}(\emph{a,d,g,j}). The intensity of this spectral component grows with $Re_\tau$ as seen from the 0.2 contour spanning a wider range of scales from left-to-right. With the current $z_{\mathcal{L}}$ location, the peak of $\phi_{\mathcal{L}}^c$ (filled blue circles) resides at $\lambda_x \approx 10\delta$ and is situated above the geometric centre between a fixed inner- and outer-scaling position, \emph{e.g.} $z^+_{gc} \equiv \sqrt{z^+_T\cdot0.15Re_\tau} = 3.46\sqrt{Re_\tau}$ (with $z^+_T = 80$), indicated with the blue open circles. Although the emergence and scaling of the broad outer-spectral peak has been documented in the literature \citep[\emph{e.g.}][]{hutchins:2007aa,mathis:2009a,rosenberg:2013a,vallikivi:2015a}, the analysis has always been approached from a total-energy perspective (observations from the measured spectra/spectrograms, $\phi_{uu}$). Clearly, when decomposing the spectrograms into physically-relevant components, the characteristic scale and wall-normal location of a type-$\mathcal{B}$ contribution, as in Figure~\ref{fig:spperry}, may change. That is, the peak in $\phi_{\mathcal{L}}^c$ does not match the peak in the spectrogram, because $\phi_{\mathcal{W}\mathcal{L}}$ (and also $\phi_{\mathcal{W}}^i$ at these still limited $Re_\tau$) constitute an energy roll-off below the outer-spectral peak in $\phi_{uu}$. Future work may benefit from reappraising scaling trends of the peak-features in the spectrogram in terms of its sub-components. Also, whether $\phi_{\mathcal{L}}^c$ will saturate to a constant state at ultra-high $Re_\tau$ remains an open question. Clearly, $\phi_{\mathcal{L}}^c$ depends on $Re_\tau$ and $z$.

Figure~\ref{fig:Retrend2}(\emph{f}) evidences the Reynolds-number independence of $\phi_{\mathcal{W}}^i$. At the small-scale end, all spectra are in close agreement \citep[expected per the law of the wall][]{samie:2018a,ganapathisubramani:2018a}. In particular, the two highest $Re_\tau$ spectra \citep[$Re_\tau \approx 13\,000$ and $19\,300$, S18:][]{samie:2018a} are in excellent agreement with the DNS spectrum \citep[$Re_\tau \approx 2\,000$, S13:][]{sillero:2013a}, since those data correspond to fully-resolved measurements \citep[\S\,\ref{sec:Redata} and detailed in][]{samie:2017phd}. Because all spectra collapse for $k_xz \apprge 10^{-1}$, the roll-offs of the experimental spectra at the low wavenumber-end also collapse because of the Reynolds-number invariant $f_W$. There, the DNS spectrum has a lower amplitude than the hot-wire-based spectra. This may partially be caused by Taylor's approximation re-distributing the VLSM spectral content in temporal spectra \citep{perry:1990a,delalamo:2009a}, but, the DNS spectrum is likely inaccurate, since it is based on spatially-developing data (\S\,\ref{sec:Redata}). In any case, the exact spectral distribution of VLSM energy should be cautioned, as it may be more broadband in spatial spectra---see Figure~17 in \citet{perry:1990a} and Figure~10 in \citet{delalamo:2009a}.

\subsection{Reynolds number trend of the $\phi_{\mathcal{W}\mathcal{L}}$ spectrum}\label{sec:Remature}
\begin{figure} 
\vspace{10pt}
\centering
\includegraphics[width = 0.999\textwidth]{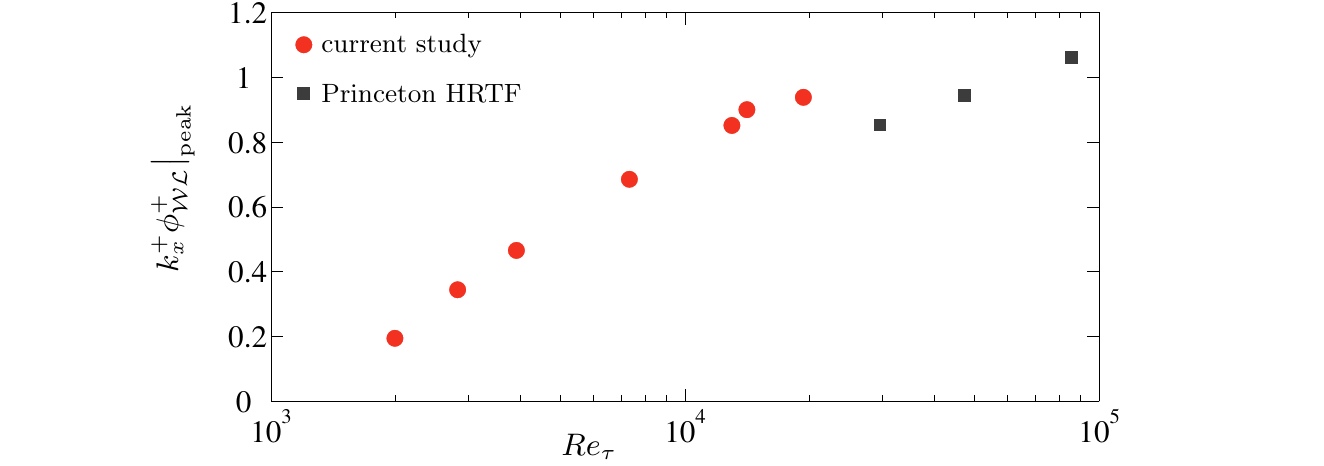}
   \caption{Peak value of the premultiplied spectral energy component $k_x^+\phi^+_{\mathcal{W}\mathcal{L}}$ at $z^+ = 100$ as function of $Re_\tau$, corresponding to the peak values of the spectra in Figures~\ref{fig:Retrend2}(\emph{c},\emph{d}). Higher Reynolds number data from the Princeton HRTF are also included \citep{vallikivi:2015a}.}
   \label{fig:maturing}
\end{figure}
With increasing $Re_\tau$, the $\phi_{\mathcal{W}\mathcal{L}}$ component grows, while the relative energy contribution of the other two spectral components, to the total energy, decays (Figure~\ref{fig:Retrend2}). At $z^+ = 100$, and at the highest $Re_\tau$ in Figure~\ref{fig:Retrend2}(\emph{d}), the spectral roll-offs of $\phi_{\mathcal{W}\mathcal{L}}$ at the large- and small-wavelength ends still overlap at $\lambda_x = 0.15R'\delta$ (recall the discussion of Figure~\ref{fig:k1pres}), meaning that $\phi_{\mathcal{W}\mathcal{L}}$ has not yet matured to a $k_x^{-1}$ region (if at all present). Our results are consistent with those of \citet{chandran:2017a}, who demonstrated that a $k_x^{-1}$ is not expected for $Re_\tau \apprle 80\,000$. Nevertheless, in order to quantify the growth of the $\phi_{\mathcal{W}\mathcal{L}}$ spectra in Figure~\ref{fig:Retrend2}(\emph{d}), their peak values are plotted in Figure~\ref{fig:maturing} as a function of $Re_\tau$. Seven peak values are plotted from the data of the current study, up to $Re_\tau \approx 19\,300$. For inspecting how the growth continues beyond $Re_\tau \approx 19\,300$, three data points are included from the Princeton High Reynolds number Test Facility (HRTF). Streamwise velocity spectra in a nominally ZPG TBL were presented by \citet{vallikivi:2015a} and from low-to-high $Re_\tau$, the three spectra were taken from data at $z^+ = 90$, 105 and 98 (close to $z^+ = 100$ for a direct comparison to the other data). The raw spectra of \citet{vallikivi:2015a} were filtered via the same filtering procedure as used in generating the $\phi_{\mathcal{W}\mathcal{L}}$ spectra in Figure~\ref{fig:Retrend2}(\emph{d}), meaning that the raw spectra were multiplied by $(f_\mathcal{W} - f_\mathcal{L}$) following formulation (\ref{eq:trdecom3}). Subsequently, its peak value was determined and plotted in Figure~\ref{fig:maturing}. The same filtering procedures ensures that, given the assumptions made in this work, the large-scale contribution $\phi_\mathcal{L}$ and small-scale portion $\phi_\mathcal{W}$ are excluded so that we reveal the scaling trend of the spectral component that obeys inner- and outer-scalings. From all data points in Figure~\ref{fig:maturing} it remains inconclusive if the peak in $\phi_{\mathcal{W}\mathcal{L}}$ plateaus to a constant value at ultra high $Re_\tau$. Our current standing with the limited data---\emph{e.g.} subject to the challenges associated with the acquisition of repeatable statistics with miniature hot-wire probes \citep[\emph{e.g.}][]{samie:2018a}---highlights the ongoing need for additional high-fidelity spectral data spanning several orders of magnitude in $Re_\tau$. Only those kind of data can provide definitive answers on whether a distinguished $k_x^{-1}$ region develops (in which case the points in Figure~\ref{fig:maturing} would plateau to a constant). Finally, it is worth noting that Figure~\ref{fig:maturing} presents the best case scenario that we have at our disposal with the available data, namely, the case of $z^+ = 100$. If a $k_x^{-1}$ region does exist for the turbulence residing at the inertial scales, it would first be visible at a location close to the wall, as that location comprises the largest range of energetic spectral scales between a $z$-scaling and outer-scaling. Here we presume that viscous effects only affect the inertial scales at a wall-normal range below a location fixed in viscous scaling (\S\,\ref{sec:kmin1}), which seems reasonable given that the inner-peak is confined to roughly $z^+ < 80$ (\emph{e.g.} from observation in Figures~\ref{fig:Retrend1}(\emph{c,f,i,l}). Following the criterion from above that $Re_\tau \apprge 80\,000$ for examining a $k_x^{-1}$ at $z^+ = 100$, in combination with the inner- and outer-scalings, it simply follows that $Re_\tau \sim \mathcal{O}(10^6)$ is required for potentially observing a $k_x^{-1}$ at $z^+ = 1000$.

\section{Concluding remarks}\label{sec:concl}
Data-driven filters for a triple decomposition of the streamwise velocity energy spectra were derived from two-point data via spectral coherence analyses. Our study addresses the need for novel decomposition techniques \citep{marusic:2017a} to fully appreciate (1) the inner-scaled, universal portion of wall-bounded turbulence, (2) a portion that would scale via the classical $k_x^{-1}$ scaling at ultra high $Re_\tau$ and which is consistent with the concept of Townsend's attached-eddies and (3) turbulence reflecting the emergence of VLSMs/superstructures with Reynolds number. Typically, spectral scalings have been researched from unaltered energy spectra alone. The current work offers new ways of conceptualizing wall-turbulence spectra by attempting to unravel the signature of classical Kolmogorov- and viscosity-dominated turbulence, co-existing with $Re_\tau$-dependent contributions. Our conclusions are listed as follows.\\[-8pt]
\begin{enumerate}[labelwidth=0.65cm,labelindent=0pt,leftmargin=0.65cm,label=(\roman*),align=left]
\item \noindent From data of $u$, spanning three decades in $Re_\tau \sim \mathcal{O}(10^3) - \mathcal{O}(10^6)$, two universal spectral filters were found. Filters $f_{\mathcal{W}}$ and $f_{\mathcal{L}}$ allow for the computation of wall-detached and wall-attached energy portions. A primary characteristic of these filters is that wavelengths smaller than $\lambda_x \approx 14z$ are, in a stochastic sense, wall-detached.\\[-10pt]
\item \noindent Filters $f_{\mathcal{W}}$ and $f_{\mathcal{L}}$ defined a spectral decomposition, introduced in \S\,\ref{sec:trdecom}. It allows a separation of $\phi_{uu}$ into an energy fraction that is wall-detached, an energy fraction that is wall-attached but does not involve global modes that reach beyond a reference $z_{\mathcal{L}}$ (typically the edge of the logarithmic region: $z_{\mathcal{L}} = 0.15\delta$) and an energy fraction that is both wall-attached and coherent with $z_{\mathcal{L}}$ (thus representing global-type modes and VLSMs).\\[-10pt]
\item \noindent Per the decomposition framework, the component representing self-similar wall-attached eddies, $\phi_{\mathcal{W}\mathcal{L}}$, contains more energy with increasing $Re_\tau$. Its peak value in the spectrum may level off (Figure~\ref{fig:maturing}), but only high-fidelity data at $z^+ = 100$ and $Re_\tau \apprge 80\,000$ (and practically $Re_\tau \apprge 10^6$, see \S\,\ref{sec:kmin1}) can provide a definite answer. These data are currently non-existent.\\[-10pt]
\item \noindent A broad spectral peak in the streamwise energy spectrogram representing VLSMs/superstructures is present even at low $Re_\tau$, when only $\phi^c_{\mathcal{L}}$ is considered. This peak resides at $\lambda_x \approx 10\delta$ (\S\,\ref{sec:Respectra}) over the range of $Re_\tau$ investigated and its amplitude appears to be a weak function of Reynolds number: perhaps the superstructure energy-trend with $Re_\tau$ is less significant than previously thought \citep{hutchins:2007aa,vallikivi:2015a,vallikivi:2015a2}. That is, in unaltered spectra the broad spectral peak exhibits a significant $Re_\tau$ dependence, but this work suggests that this is an artifact of the growing $\phi_{\mathcal{W}\mathcal{L}}$ component.\\[-8pt]
\end{enumerate}
\begin{figure} 
\vspace{10pt}
\centering
\includegraphics[width = 0.999\textwidth]{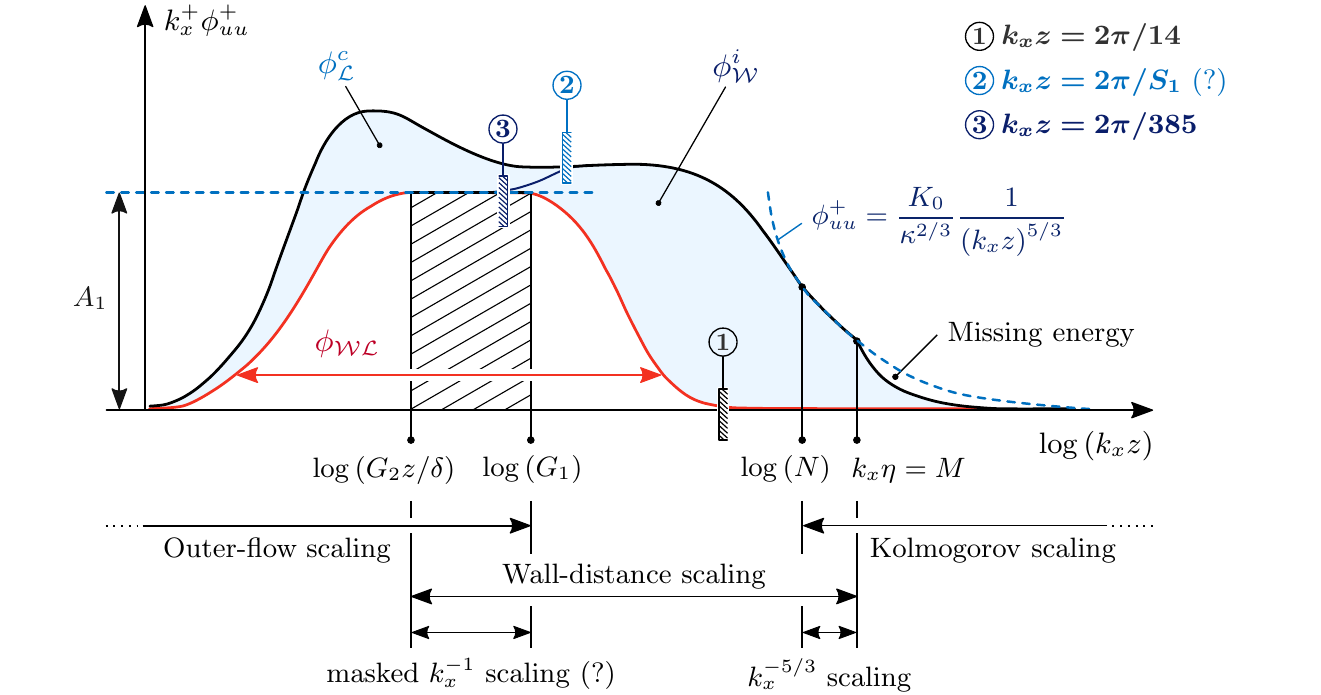}
   \caption{Structure of $k_x\phi_{uu}$ within the logarithmic region of TBLs, following \citet{perry:1977a} and \citet{perry:1986a} and reproduced from \citet{marusic:1995a} and updated following the current findings from the triple decomposition technique based on $u$ fluctuations alone at $Re_\tau$ up to $\mathcal{O}(10^4)$.}
   \label{fig:concl}
\end{figure}
Figure~\ref{fig:concl} summarizes concrete findings of spectral sub-components forming the streamwise energy spectra in ZPG TBL flow (following the hypothesized structure in Figure~\ref{fig:spperry}). In the context of our limitations of dealing with quantity $u$ only (and the stochastic decomposition approach), we can conclude that $\phi_{\mathcal{W}\mathcal{L}}$ is constrained at $k_xz < 2\pi/R \approx 2\pi/14 \approx 0.45$ (boundary 1). Wall-incoherent energy $\phi^i_{\mathcal{W}}$ is bounded by $k_xz = 2\pi/R/\exp(1/C_1) \approx 2\pi/385 \approx 0.016$ at high $Re_\tau$ (boundary 3). Finally, component $\phi^c_{\mathcal{L}}$, comprising VLSM and global mode-energy, overlaps with the two other components at $Re_\tau$ up to at least $\mathcal{O}(10^5)$. Its high wavenumber-limit can obey different scalings, which can only be addressed with future research: high Reynolds number data with converged large-scale energy content are crucial, in combination with a novel spectral decomposition that is able to unambiguously separate wall-attached self-similar turbulence from non-self-similar contributions. When the high wavenumber-end would be fixed in $z$ scaling ($S_1$ of boundary 2 would be a constant), a potential $k_x^{-1}$ scaling may always remain obscured by other energy components. Only when this boundary scales with $\delta$ (or a combination of $\delta$ and $z$), $S_1$ would constitute a $Re_\tau$ dependence and ultra high $Re_\tau$ data may start to expose a $k_x^{-1}$ scaling in unaltered, measured spectra.

In Part 2 \citep{baars:part2}, scaling trends of the integrated energy spectra (the turbulence intensity) are considered in both the outer- and near-wall regions of the flow.

\section*{Acknowledgements}
The authors wish to gratefully acknowledge the Australian Research Council for financial support and are appreciative of the publicly available DNS data of \citet*{sillero:2013a}. We would also like to give special thanks to Dougal Squire for experimental support and to Jason Monty, Dominik Krug, Dileep Chandran and Hassan Nagib for helpful discussions on the content of the manuscript.

\appendix
\section{Additional figures}\label{sec:appA}
\subsection{Energy spectra with logarithmic ordinate}\label{sec:appA1}
\begin{figure} 
\vspace{10pt}
\centering
\includegraphics[width = 0.999\textwidth]{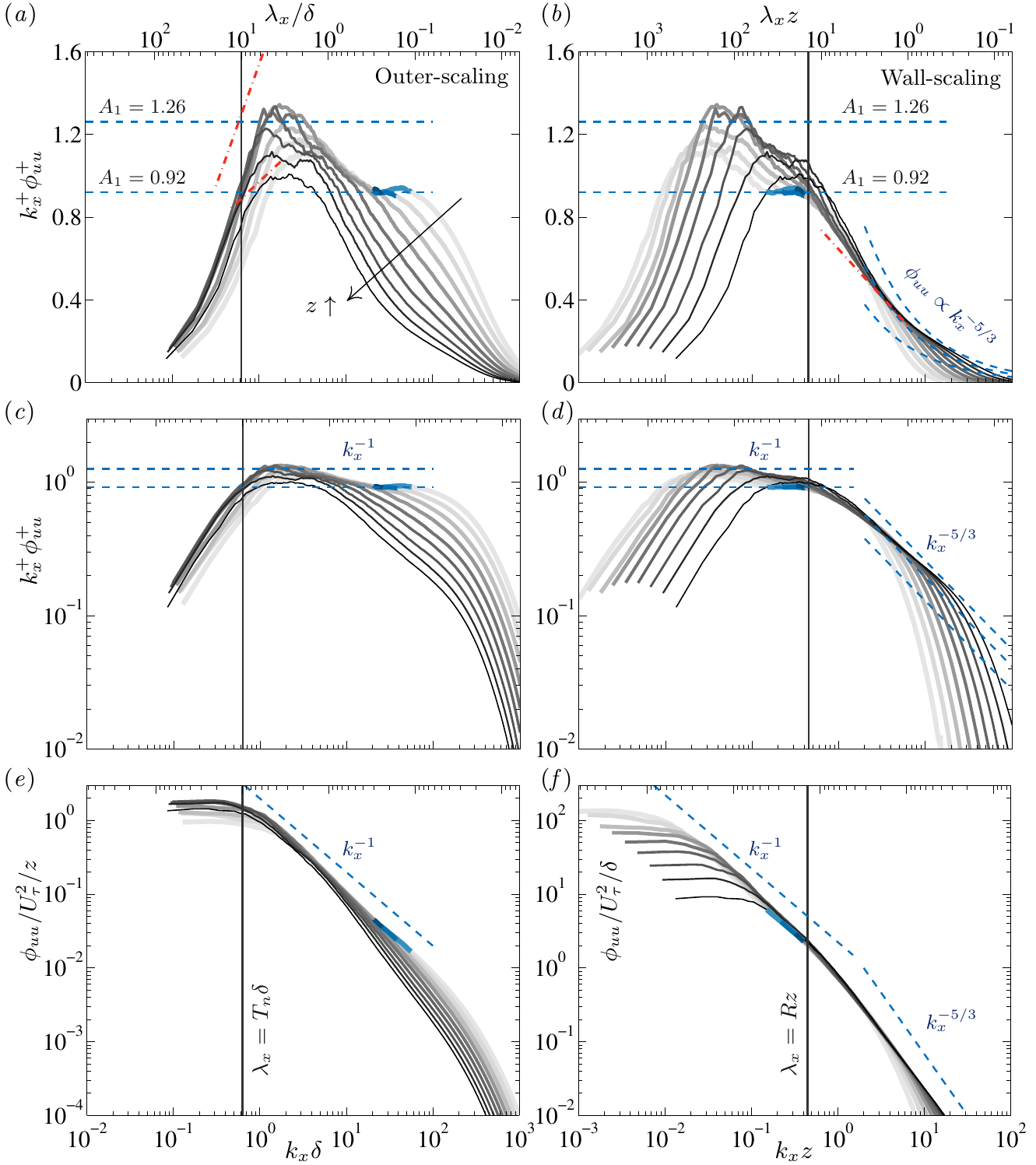}
   \caption{Sequence of energy spectra from Figure~\ref{fig:spintro2}, presented with different ordinates. (\emph{a},\emph{b}) Premultiplied spectra with a linear ordinate. (\emph{c},\emph{d}) Premultiplied spectra with a logarithmic ordinate. (\emph{e},\emph{f}) Regular spectra with a linear ordinate. Sub-figures in the left and right columns comprise an abscissae in outer-scaling and wall-scaling, respectively. Note: $\lambda_x \equiv U(z)/f$.}
   \label{fig:spappA}
\end{figure}
The sequence of energy spectra shown in Figure~\ref{fig:spintro2} are re-shown in Figures~\ref{fig:spappA}(\emph{a},\emph{b}). These spectra at nine logarithmically-spaced wall-normal positions in the range $100 \apprle z^+ \apprle 0.15\delta^+$ at $Re_\tau \approx 14\,100$ are compared to the spectral scaling proposed by \citet{delalamo:2004a}. At the large-scale end (Figure~\ref{fig:spappA}\emph{a}) the red dash-dot lines indicate (3.3) of \citet{delalamo:2004a}, here taken as $k^+_x\phi^+_{uu} = f(z/\delta)\log\left(24 \delta^2 k_x\delta/(2\pi z \delta)\right)$, with $f \approx 6.5\times 10^{-4}$ for the highest $z$ position-spectrum ($z/\delta \approx 0.15$) and $f \approx 2.0\times 10^{-4}$ for the lowest $z$ position ($z/\delta \approx 0.0072$). The two bounding trends are shown for $0.26 < k_x\delta < 1.04$. The large-scale motion energy in the TBL spectra appears to be less than what is predicted from the aforementioned relation. This could be a Reynolds number effect, or a difference in the VLSM content between TBL and channel flow (the formulation was found from lower Reynolds number channel data). At the small-scale ends of the spectra (Figure~\ref{fig:spappA}\emph{b}), relation (3.2) of \citet{delalamo:2004a} is compared to the data: $k^+_x\phi^+_{uu} = \beta \log\left(2\pi\alpha^2/(k_xz)\right)$ with $\alpha = 2$ and $\beta = 0.2$ over a range $0.63 < k_xz < 6.3$. The TBL spectra show a different slope \citep[as was also noted by][]{vallikivi:2015a}. Empirical scalings for spectra dependent on the flow type (channel, TBL, etc.), vary with $Re_\tau$ and depend on $z$. Hence, scalings cannot be extended to high-Reynolds-number data.

For reference, the sequence of spectra in Figures~\ref{fig:spappA}(\emph{a},\emph{b}) are shown in Figures~\ref{fig:spappA}(\emph{c},\emph{d}) and Figures~\ref{fig:spappA}(\emph{e},\emph{f}) with different ordinates. For broadband turbulence, the interpretation of energy content in the spectra is promoted by recognizing that
\begin{eqnarray}
 \label{eq:intspec}
 \overline{u^2}^+\left(z\right) = \int \phi^+_{uu}\left(z;k_x\right){\rm d}k^+_x = \int k^+_x\phi^+_{uu}\left(z;k_x\right){\rm d}\ln \left(k^+_x\right),
\end{eqnarray}
where $\overline{u^2}$ is the variance of $u$. Thus, premultiplied spectra with a linear ordinate and a logarithmic abscissae represent an area that is proportional to fluctuation energy (and naturally enhances the energy containing range). When presented with a logarithmic ordinate (Figures~\ref{fig:spappA}\emph{c},\emph{d}), and without premultiplication (Figures~\ref{fig:spappA}\emph{e},\emph{f}), it may visually enhance the appearance of a $\phi_{uu} \propto k_x^{-1}$ region (similarly for the $k^{-5/3}$ region). For scrutinization of the energy containing range and the examination of the $k_x^{-1}$ it thus preferred to present premultiplied spectra as in Figures~\ref{fig:spappA}(\emph{a},\emph{b}).

\subsection{Coherence spectrograms with respect to logarithmic-region positions}\label{sec:appA2}
\begin{figure} 
\vspace{10pt}
\centering
\includegraphics[width = 0.999\textwidth]{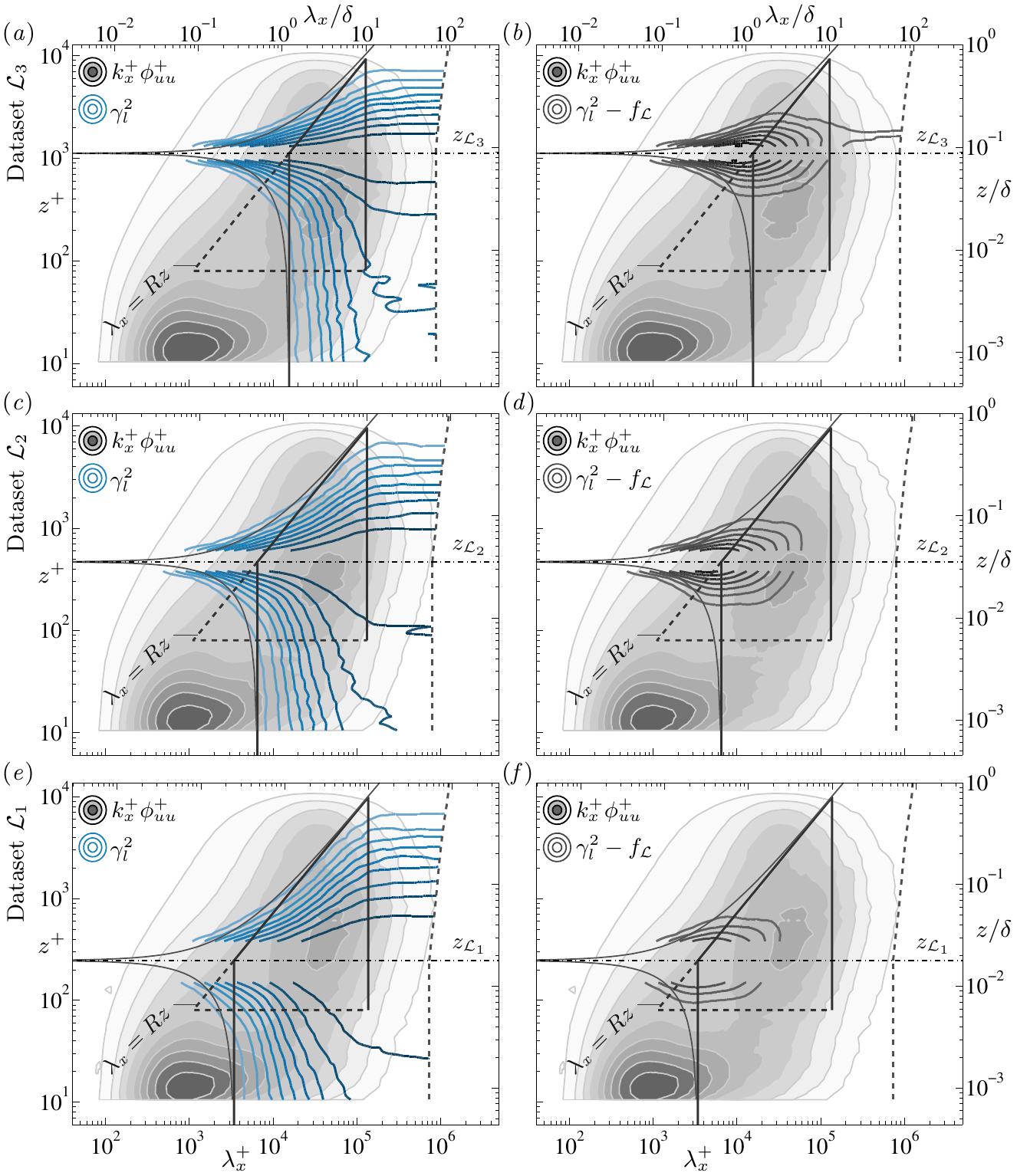}
   \caption{(\emph{a},\emph{c},\emph{e}) Similar to Figure~\ref{fig:LCSdataL} but for all three datasets $\mathcal{L}_{1-3}$. Coherence spectrograms $\gamma^2_l(z,z_{\mathcal{L}};\lambda_x)$, relative to reference $z^+_{\mathcal{L}}$ (iso-contours 0.1:0.1:0.9), superposed on its associated premultiplied energy spectrogram $k^+_x\phi^+_{uu}$ (filled iso-contours 0.2:0.2:1.8). (\emph{b},\emph{d},\emph{f}) Coherence spectrograms $\gamma^2_l(z,z_{\mathcal{L}};\lambda_x)$ with the $f_{\mathcal{L}}$ filter subtracted, representing wall-detached turbulence that is $z_{\mathcal{L}}$-coherent. Note: $\lambda_x \equiv U(z)/f$.}
   \label{fig:LCSappA}
\end{figure}

Figure~\ref{fig:LCSdataL} presents the coherence spectrogram for dataset $\mathcal{L}_3$ only. Here, in Figures~\ref{fig:LCSappA}(\emph{a},\emph{c},\emph{e}), contours of $\gamma^2_l(z,z_{\mathcal{L}};\lambda_x)$ are shown for datasets $\mathcal{L}_{1-3}$. With varying $\mathcal{L}$, the general trend as discussed in \S\,\ref{sec:logcoh}, remains valid. That is, for $z \ll z_{\mathcal{L}}$, say $z < z_{\mathcal{L}}/8$, the coherence spectra become non-zero for $\lambda_x \apprge Rz_{\mathcal{L}}$ (the vertical line extending down to the abscissa) and this is representative of the wall-attached motions with an extent $z < z_{\mathcal{L}}$ not being coherent with reference location $z < z_{\mathcal{L}}$. For $z \gg z_{\mathcal{L}}$, coherence spectra tend towards the same trend as in Figure~\ref{fig:LCSdataW}.

When $z \sim z_{\mathcal{L}}$, coherence spectra in Figures~\ref{fig:LCSappA}(\emph{a},\emph{c},\emph{e}) reveal that even scales smaller than $\lambda_x < Rz$ become coherent with $z_{\mathcal{L}}$. This implies that turbulent scales for which $\gamma^2_l(z,z_{\mathcal{L}};\lambda_x) > \gamma^2_l(z,z_{\mathcal{W}};\lambda_x)$ are wall-detached scales that are $z_{\mathcal{L}}$-coherent. Figures~\ref{fig:LCSappA}(\emph{b},\emph{d},\emph{f}) present the $\gamma^2_l(z,z_{\mathcal{L}};\lambda_x)$ spectrograms with the $f_{\mathcal{L}}$ filter subtracted and thus indicate the energy portion of the spectra that is $z_{\mathcal{L}}$-coherent and $z_{\mathcal{W}}$-incoherent. The iso-contours centred around $z_{\mathcal{L}}$ are widest (spanning the largest range in $z$) around $\lambda_x \approx Rz$, indicating the trivial fact that the smallest scales can only be coherent for very small separations $\Delta z$ (wall-normal distance $\Delta z$ is schematically shown in Figure~\ref{fig:expsetup}). Lines of $\lambda_x/\Delta z = R$ are drawn in all sub-figures of Figure~\ref{fig:LCSappA} (the black, curved lines, approaching $z = z_\mathcal{L}$ from both below/above with decreasing $\lambda_x$). If $\gamma^2_l$ iso-contours follow these lines, the aforementioned wall-detached scales may still comprise the same self-similar aspect ratio ($R \approx 14$) as the wall-attached turbulence \citep[imagine detached turbulence that might be remnants of eddies once attached earlier in their lifetimes and reflecting Townsend's AEH,][]{marusic:2019a}. Generally, $\gamma^2_l$ iso-contours drift towards smaller wavelengths for $z \rightarrow z_{\mathcal{L}}$: this is expected as $\gamma^2_l \rightarrow 1$ for all $\lambda_x$ at $z = z_{\mathcal{L}}$. A further detailed examination of the coherence spectra in Figures~\ref{fig:LCSappA}(\emph{b},\emph{d},\emph{f}) is outside the scope of this work.

\subsection{Energy spectra at $z^+ = 100$ with outer-scaling and wall-scaling}\label{sec:appA3}
\begin{figure} 
\vspace{10pt}
\centering
\includegraphics[width = 0.999\textwidth]{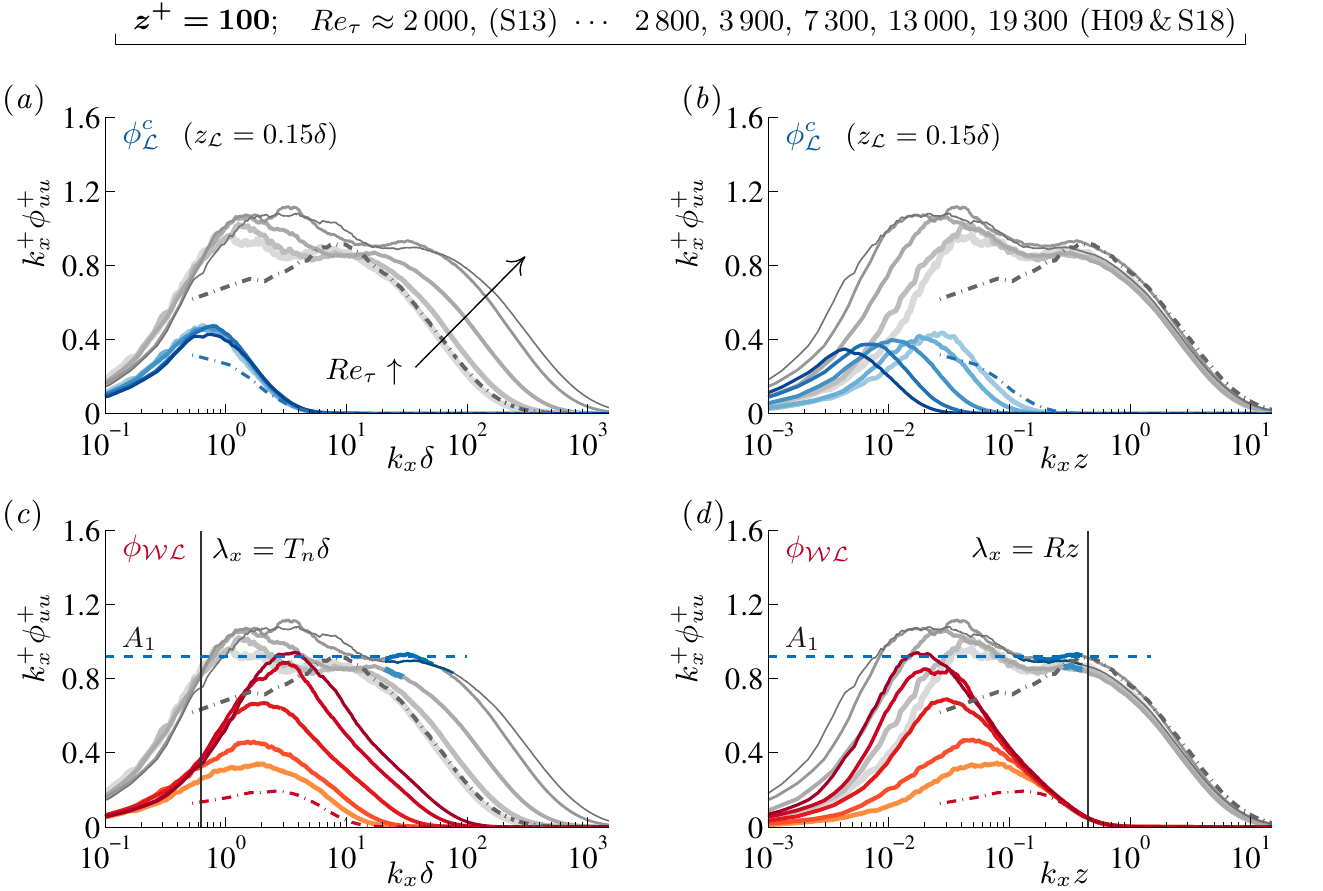}
   \caption{Similar to Figures~\ref{fig:Retrend2}(\emph{a}-\emph{d}) but with the outer-scaling abscissae in (\emph{a},\emph{c}) employing $k_x \equiv 2\pi f/U(z_{\mathcal{L}})$, while the abscissae in (\emph{b},\emph{d}) are indifferent, employing $k_x \equiv 2\pi f/U(z)$. Premultiplied energy spectra at $z^+ = 100$ for six values of $Re_\tau$. Total energy spectra are shown in each sub-figure with light grey. In each of the three rows, spectra corresponding to one specific spectral sub-component are superposed. Data are from S13: \citet{sillero:2013a}, H09: \citet{hutchins:2009a} and S18: \citet{samie:2018a}.}
   \label{fig:Retrend2App}
\end{figure}

A Reynolds number trend appears at the large-scale end of the spectra in Figures~\ref{fig:Retrend2}(\emph{a},\emph{c}). In Figures~\ref{fig:Retrend2App}(\emph{a},\emph{c}) the local mean velocity at $z_{\mathcal{L}} = 0.15\delta$ (as opposed to the local mean velocity at $z^+ = 100$) is utilized in constructing the scale-axis. Now the large-scale end of the spectra exhibit a better collapse. Since these large-scale motions \citep[or global modes,][]{delalamo:2003a} are expected to convect with a velocity that is set by an outer scale, the collapse suggests that the energy content in these motions may be Reynolds-number invariant, at least over the range of $Re_\tau$ considered here.

\bibliographystyle{jfm}
\bibliography{bibtex_database}

\end{document}